\documentclass[12pt]{article}
\usepackage{graphicx}
\usepackage{amsfonts}
\usepackage{amssymb,amsmath}
\usepackage{color}
\usepackage{cite}
\usepackage[colorlinks=true,linkcolor=blue,citecolor=blue]{hyperref}

\begin{document}

\begin{titlepage}

\begin{flushright}
IPMU 12-0233\\
ICRR-Report-642-2012-31
\end{flushright}

\begin{center}

{\Large \bf  
Extended analysis of 
CMB constraints on non-Gaussianity in
isocurvature perturbations
}

\vskip .45in

{
Chiaki Hikage$^{a}$,
Masahiro Kawasaki$^{b,c}$,
Toyokazu Sekiguchi$^d$ and
Tomo Takahashi$^e$
}

\vskip .45in

{\em
$^a$Kobayashi-Maskawa Institute, Nagoya University, Nagoya 464-8602, Japan \vspace{0.2cm}\\
$^b$Institute for Cosmic Ray Research,
University of Tokyo, Kashiwa 277-8582, Japan \vspace{0.2cm}\\
$^c$ Kavli Institute for the Physics and Mathematics of the Universe,
University of Tokyo, Kashiwa 277-8568, Japan \vspace{0.2cm}\\
$^d$Department of Physics and Astrophysics, Nagoya University, Nagoya 464-8602, Japan\\
$^e$Department of Physics, Saga University, Saga 840-8502, Japan
}

\end{center}

\vskip .4in

\begin{abstract}

We study CMB constraints on non-Gaussianity from isocurvature 
perturbations of general types.  Specifically, 
we study CDM/neutrino isocurvature perturbations
which are uncorrelated or totally correlated with adiabatic ones.
Using the data from the WMAP 7-year observation at V and W bands, we obtained
optimal constraints on the nonlinearity parameters of adiabatic and
isocurvature perturbations. Our result shows that primordial perturbations
are consistent with Gaussian ones at around 2$\sigma$ level for 
above mentioned isocurvature modes.

\end{abstract}

\end{titlepage}

\setcounter{page}{1}

\section{Introduction} 
\label{sec:introduction}

The adiabaticity of primordial density perturbations are one of the most important probes 
to understand the early Universe.
Although current cosmological observations suggest that primordial fluctuations are 
almost adiabatic and the contribution from isocurvature fluctuations is  severely constrained, 
 once any deviation from the adiabatic fluctuations are 
detected even if it is tiny, it readily indicates that there are multiple sources of 
density fluctuations. 

 In the view of its importance, there have been many works on 
constraints on isocurvature fluctuations by using power spectrum \cite{Moodley:2004nz,Beltran:2004uv,Bean:2006qz,Trotta:2006ww,Keskitalo:2006qv,Kawasaki:2007mb,Sollom:2009vd,Valiviita:2009bp,Li:2010yb,
Valiviita:2012ub}. 
However, current cosmological observations are so precise that 
its non-Gaussian nature  can also be measured/constrained  
in addition to the power spectrum\footnote{
Future observations of redshifted 21 cm line emission can 
also probe the isocurvature perturbations 
\cite{Gordon:2009wx,Kawasaki:2011ze}. 
}.

Compared to the counterpart of adiabatic fluctuations, observational constraints on isocurvature 
non-Gaussianity using actual data have almost not been studied\footnote{
See Refs. \cite{Hikage:2009rt,Langlois:2011hn,Kawakami:2012ke,Langlois:2012tm} for expected constraints from future observations by using Fisher matrix analysis. For theoretical aspects of isocurvature non-Gaussianity, see Refs. \cite{Kawasaki:2008sn,Langlois:2008vk,Kawasaki:2008pa,Kawakami:2009iu,Takahashi:2009cx,Langlois:2010fe,Kawakami:2012ke,Langlois:2011zz}. 
}, 
except a few works \cite{Hikage:2008sk,Hikage:2012be}. 
In Ref. \cite{Hikage:2012be}, we have obtained observational constraints on isocurvature non-Gaussianity
by adopting an optimal bispectrum estimator, focusing on 
uncorrelated type of CDM (baryon) mode. The constraints
would give interesting implications to particle physics as well 
since this type of  
isocurvature fluctuations can be 
generated in some models such as the axion model 
\cite{Axenides:1983hj,Seckel:1985tj,Linde:1985yf,
Linde:1990yj,Turner:1990uz,Linde:1991km,Lyth:1991ub,Kasuya:2009up} 
and Affleck-Dine baryogenesis 
\cite{Enqvist:1998pf,Enqvist:1999hv,Kawasaki:2001in}.

In fact,  isocurvature fluctuations can also be 
correlated with adiabatic ones in some cases and such correlated type is naturally produced
in the curvaton model \cite{Enqvist:2001zp,Lyth:2001nq,Moroi:2001ct}, 
which has been attracting attention due to its possibility of getting large non-Gaussianity. 
In the model,  depending on when and how DM or baryon asymmetry are produced, 
positively and negatively correlated isocurvature fluctuations 
can be generated \cite{Moroi:2002rd,Lyth:2002my,Moroi:2002vx,Lyth:2003ip,Hamaguchi:2003dc,Gordon:2003hw,Ikegami:2004ve,Ferrer:2004nv,Beltran:2008ei,Moroi:2008nn,Lemoine:2009yu,Lemoine:2009is,Takahashi:2009cx}.    Furthermore, isocurvature fluctuations can also be associated with
neutrino  \cite{Bucher:1999re}, which generates neutrino density isocurvature mode\footnote{
There is also another isocurvature mode associated with neutrino
called neutrino isocurvature velocity (NIV) mode \cite{Bucher:1999re}, 
in which there is difference between initial velocity perturbations 
of neutrino and photon. However, there are so far no concrete models
which can generate NIV perturbations at super-horizon scales. We do not study NIV mode  in this paper.
}. 
Such a neutrino mode can be related to lepton asymmetry, thus its constraints may give interesting implications to the lepton sector. 

In the light of these considerations, 
it would be also worth investigating observational constraints on isocurvature non-Gaussianities 
of correlated type and/or other modes in addition to CDM (baryon) one such as the neutrino mode
mentioned above. 
The main purpose of this paper is to study the constraints on non-Gaussianity of those 
kinds of isocurvature types/modes by
extending the analysis of our previous study \cite{Hikage:2012be}. 
By using the optimal estimator which we adopted there,
we investigate observational constraints on non-linearity parameters 
for correlated/uncorrelated CDM and neutrino 
isocurvature modes from WMAP7 data. Implications of our
results are also discussed for some explicit model.

The organization of this paper is as follows. In Section \ref{sec:NG}, 
we present the models of non-Gaussian isocurvature perturbations
which we are to constrain in this paper and their CMB bispectra. 
We also make the Fisher matrix analysis to get some idea for the 
size of the error, which also serves as a check of our results presented in Section 
\ref{sec:result}. 
Implications of our constraints are discussed for the curvaton scenario in Section \ref{sec:curvaton}.
The final section is devoted to conclusion.

In this paper, we denote the mean value and perturbation of 
a fluctuating field by bar and delta, respectively. 
A bracket $\langle\,\rangle$ represents an ensemble average.

\section{Non-Gaussian primordial perturbations and CMB bispectrum} 
\label{sec:NG}

\subsection{Non-Gaussian perturbations}
\label{sec:primordial}

Following mostly Refs. \cite{Kawasaki:2008pa,Kawasaki:2008sn}, 
here we briefly summarize
the formalism to discuss non-Gaussianities of primordial 
curvature and isocurvature perturbations.
Based on the $\delta N$-formalism \cite{Sasaki:1995aw,Lyth:2004gb,
Lyth:2005du,Lyth:2005fi}, the curvature perturbation
on the uniform density hypersurface of a component $i$, 
which we denote $\zeta_i$, can be given by
\begin{equation}
\zeta_i(\vec x,t)=N_i(\vec x, t)-\bar N_i(t),
\end{equation}
where $N(\vec x,t)$ is an $e$-folding number from 
the initial flat hypersurface
to the uniform density hypersurface  of $i$ at time $t$. 
In particular, we denote the curvature perturbation
on the uniform density hypersurface of the total matter 
as $\zeta_{\rm tot}$. 
Then we adopt the following definition of the 
curvature $\zeta$ and isocurvature perturbations $S_i$, 
\begin{eqnarray}
\zeta(\vec x,t)&\equiv&\zeta_{\rm tot}(\vec x,t), \\
S_i(\vec x,t)&\equiv&3\{\zeta_i(\vec x,t)-\zeta_{\rm tot}(\vec x,t)\}.
\end{eqnarray}
In this paper, we consider two types of primordial isocurvature perturbations, 
which are CDM one (CI) and neutrino density one (NID).

In the picture of the separate Universe, the spatial dependence of 
$N_i(\vec x,t)$ should come only from the initial values of 
the degrees of freedom (or scalar fields) during inflation, which we write as
$\phi_a(\vec x)=
\bar\phi_a+\delta\phi_a(\vec x)$ 
with a subscript $a$ represents a species of scalar fields.
We can expand the curvature and isocurvature perturbations $\zeta$ 
and $S_i$ locally in terms of $\phi$ up to second order, which leads to
\begin{eqnarray}
\zeta(\vec x,t)&=&N_{{\rm tot},a}(t)\delta\phi_a(\vec x)+\frac12N_{{\rm tot},ab}(t)
\{\delta\phi_a(\vec x)\delta\phi_b(\vec x)-
\langle\delta\phi_a(\vec x)\delta\phi_b(\vec x)\rangle\}+\cdots, \\
S_i(\vec x,t)&=&S_{i,a}(t)\delta\phi_a(\vec x)+\frac12S_{i,ab}(t)
\{\delta\phi_a(\vec x)\delta\phi_b(\vec x)-
\langle\delta\phi_a(\vec x)\delta\phi_b(\vec x)\rangle\}+\cdots, 
\end{eqnarray}
where the coefficients with subscripts $a$, $b$, etc. attached to $N$ represent 
derivatives with respect to the initial fields, e.g., 
\begin{equation}
N_{{\rm tot},a}(t)=
\left.
\frac{\partial N_{\rm tot}(\vec x,t)}
{\partial \phi_a(\vec x)}\right|_{\phi(\vec x)=\bar\phi}.
\end{equation}

In this paper, we assume that the initial fluctuations $\delta\phi_a$ 
are Gaussian and different initial fields are
uncorrelated, i.e. $\langle \delta\phi_a\delta\phi_b\rangle=0$ for
$a\ne b$. 
Nonlinearity parameters of the local type for the curvature and isocurvature perturbations are defined as follows:
\begin{eqnarray}
\zeta(\vec x)&=&\zeta_{\rm G}(\vec x)
+\frac35f_{\rm NL}(\zeta_{\rm G}^2(\vec x)-\langle\zeta_{\rm G}^2(\vec x)\rangle), 
\label{eq:zeta} \\
S_i(\vec x)&=&S_{i{\rm G}}(\vec x)
+f^{\rm (ISO)}_{\rm NL}(S_{i{\rm G}}^2(\vec x)-\langle S_{i{\rm G}}^2(\vec x)\rangle),
\label{eq:S} 
\end{eqnarray}
where $\zeta_{\rm G}$ and $S_{i{\rm G}}$ are the Gaussian part of $\zeta$ and $S_i$, respectively.
Eqs. \eqref{eq:zeta} and \eqref{eq:S} can be recast into a general form:
\begin{equation}
X^A(\vec x)=X^A_{\rm G}(\vec x)
+f_{\rm NL}^A(X^A_{\rm G}(\vec x)^2-\langle X^A_{\rm G}(\vec x)^2\rangle), 
\label{eq:X}
\end{equation}
where $X_{\rm G}^A$ is the Gaussian part of $X^A$.
Here we defined a  non-linearity parameter $f_{\rm NL}^A$,
which is related to the adiabatic and isocurvature ones via 
$f_{\rm NL}^\zeta=\frac35f_{\rm NL}$ and $f_{\rm NL}^S=f_{\rm NL}^{\rm (ISO)}$.

We present constraints in two limiting cases, both of which are of particular interest.
In one limiting case, $\zeta_{\rm G}$ and $S_{i{\rm G}}$ are 
assumed to be uncorrelated, i.e. $\langle\zeta_{\rm G}S_{i{\rm G}}\rangle=0$.
This is the case that we discussed for CDM (baryon) isocurvature mode in the previous paper \cite{Hikage:2012be}.
In terms of the scalar field fluctuations, this corresponds to the case that
the curvature and isocurvature perturbations are generated from two different fields 
$\phi$ and $\sigma$, i.e. $\zeta_{\rm G}\propto\delta \phi$
and $S_{i{\rm G}}\propto\delta \sigma$.
As we have stated in Introduction, this form of non-Gaussian
isocurvature perturbations can be generated in e.g. the axion model \cite{Axenides:1983hj,Seckel:1985tj,Linde:1985yf,
Linde:1990yj,Turner:1990uz,Linde:1991km,Lyth:1991ub,Kasuya:2009up}
and the Affleck-Dine mechanism \cite{Enqvist:1998pf,Enqvist:1999hv,Kawasaki:2001in}.
In another limiting case, $\zeta_{\rm G}$ and $S_{i{\rm G}}$ are assumed to be 
totally (anti-) correlated, so that $\langle\zeta_{\rm G}S_{i{\rm G}}\rangle=\pm \sqrt{
\langle\zeta_{\rm G}\zeta_{\rm G}\rangle\langle S_{i{\rm G}}S_{i{\rm G}}\rangle}$.
This corresponds to the cases where $\zeta$ and $S_{i{\rm G}}$ are
generated from only a single field $\phi$, so that $\zeta_{\rm G}\propto\
S_{i{\rm G}}\propto \delta\phi$. Such a case can be realized in e.g. the curvaton
scenario \cite{Moroi:2002rd,Lyth:2002my}.

From Eq. \eqref{eq:X}, the primordial bispectrum can be given by
\begin{equation}
\langle X^{A}(\vec k_1) X^{B}(\vec k_2)X^{C}(\vec k_3) \rangle
=[f^{A}_{\rm NL} P_{AB}(k_2)
P_{AC}(k_3)
+(\mbox{2 perms})](2\pi)^3\delta^{(3)}(\vec k_1+\vec k_2+\vec k_3),
\label{eq:primb}
\end{equation}
where $P_{AB}(k)$ is the power spectrum of (cross-)correlation
between $X^A_{\rm G}$ and $X^B_{\rm G}$, i.e.
\begin{equation}
\langle X_{\rm G}^A(\vec k)X_{\rm G}^B(\vec k^\prime)\rangle 
\equiv P_{AB}(k) (2\pi)^3\delta(\vec k+\vec k^\prime).
\label{eq:power}
\end{equation}
Note that in Eq. \eqref{eq:primb}, we omitted the loop contributions, 
assuming that the tree terms are leading ones in the power spectrum and the non-linearity parameters.

Current observations have not detected any isocurvature perturbations 
and only the upper bounds on the amplitudes of the isocurvature power spectra
have been provided. The upper bounds depend on the kinds of isocurvature modes 
and their correlations with the curvature perturbations.
We define the ratio of the amplitude of isocurvature power spectrum
to that of adiabatic one  at the reference scale $k_*$ as
\begin{equation}
\alpha\equiv \frac{P_{S_iS_i}(k_*)}{P_{\zeta\zeta}(k_*)}.
\end{equation}
In this paper, we take the WMAP convention $k_*=0.002$Mpc$^{-1}$.
In the case of CI, WMAP 7-year result \cite{Komatsu:2010fb}
gives upper bounds\footnote{
In Ref. \cite{Komatsu:2010fb}, 
only the positively-correlated CI model is studied, 
and does not investigate the negatively-correlated one.
However, since 
$\alpha$ allowed observationally is so small, that the upper bounds on $\alpha$
for totally correlated cases would not depend much on the sign of the correlation.
}
\begin{eqnarray}
\alpha&<&0.15\quad(\mbox{uncorrelated}), 
\label{eq:CI_uncorr} \\
\alpha&<&0.008\quad(\mbox{totally correlated}). 
\label{eq:CI_corr}
\end{eqnarray}
For the case of NID, a combined analysis from the WMAP 7-year 
and the ACT data \cite{Dunkley:2010ge} 
gives upper bounds \cite{Kawasaki:2011rc}\footnote{
We note that the effective number of neutrinos $N_{\rm eff}$ 
is also varied in Ref. \cite{Kawasaki:2011rc}. However,  
the degeneracy between $\alpha$ and $N_{\rm eff}$
is not severe, thus we expect that constraints with fixed $N_{\rm eff}=3$
should be almost the same. 
For most recent constraints on $\alpha$ with fixed 
$N_{\rm eff}$, we refer to Ref. \cite{Bean:2006qz}, which uses
the WMAP 3-year data. 
}
\begin{eqnarray}
\alpha&<&0.19\quad(\mbox{uncorrelated}), \\
\alpha&<&0.013\quad(\mbox{totally correlated}).
\label{eq:NID_corr}
\end{eqnarray}

Here some comments on the parametrization of non-linearity parameters 
are in order. 
When we consider the bispectrum with multiple sources, one can generally 
define the non-linearity parameters  $\tilde{f}^{A, BC}_{\rm NL}$ as follows \cite{Langlois:2011hn}:
\begin{eqnarray}
\langle X^{A}(\vec k_1)  X^{B}(\vec k_2) X^{C}(\vec k_3)  \rangle
&&=
(2\pi)^3\delta^{(3)}(\vec k_1+\vec k_2+\vec k_3)  \notag \\
 &&  \times \left[
\tilde{f}^{A, BC}_{\rm NL} P_\zeta (k_2) P_\zeta (k_3) 
+ \tilde{f}^{B, CA}_{\rm NL} P_\zeta (k_2) P_\zeta (k_3) 
+ \tilde{f}^{C, AB}_{\rm NL} P_\zeta (k_2) P_\zeta (k_3) 
\right],  \notag \\
\label{eq:general_fNL}
\end{eqnarray}
where $\tilde{f}_{\rm NL}^{A, BC} = \tilde{f}_{\rm NL}^{A, CB}$ is satisfied due to the symmetric structure for the 
latter two indices and the (total) curvature power spectrum is used instead of $P_{AB}$ defined in Eq.~\eqref{eq:power}. 
When we consider the case, where curvature and isocurvature 
perturbations (of a single type) coexist, 
there exist 6 non-linearity parameters in general: 
$ 
\tilde{f}_{\rm NL}^{\zeta, \zeta\zeta}, 
\tilde{f}_{\rm NL}^{\zeta, \zeta S}, 
\tilde{f}_{\rm NL}^{\zeta, SS},
\tilde{f}_{\rm NL}^{S, \zeta\zeta},  
\tilde{f}_{\rm NL}^{S, S \zeta}, 
\tilde{f}_{\rm NL}^{S, SS} 
$.
However, if one considers uncorrelated or totally correlated isocurvature models, 
these parameters can be reduced or related in some way. 
By using $f_{\rm NL}^\zeta = \frac35 f_{\rm NL}$ and $f_{\rm NL}^S=f_{\rm NL}^{\rm (ISO)}$ defined in Eqs.~\eqref{eq:zeta} and \eqref{eq:S}, 
for uncorrelated models, the above non-linearity parameters are given by
\begin{eqnarray}
\tilde f^{\zeta,\zeta\zeta}_{\rm NL}= f_{\rm NL}^\zeta, 
\qquad \qquad  
\tilde f^{S,SS}_{\rm NL}=\alpha^2 f_{\rm NL}^S,
\qquad \qquad  
 \tilde f^{A,BC}_{\rm NL} =0 \mbox{~~(others)}.
\end{eqnarray}
For totally correlated models, we have
\begin{eqnarray}
\tilde f^{\zeta,\zeta\zeta}_{\rm NL} =f_{\rm NL}^\zeta, 
\qquad \qquad  
\tilde f^{\zeta,\zeta S}_{\rm NL} &=& \pm \alpha^{1/2} f_{\rm NL}^\zeta, 
\qquad \qquad  
\tilde f^{\zeta,SS}_{\rm NL} =\alpha f_{\rm NL}^\zeta, \notag \\
\tilde f^{S,\zeta\zeta}_{\rm NL} =\alpha f_{\rm NL}^S, 
\qquad \qquad  
\tilde f^{S,\zeta S}_{\rm NL}  &=&\pm \alpha^{3/2} f_{\rm NL}^S, 
\qquad \qquad  
\tilde f^{S,SS}_{\rm NL} =\alpha^2 f_{\rm NL}^S, 
\label{eq:corr_fNL}
\end{eqnarray}
where the signs should be +/- if isocurvature perturbations
are positively/negatively correlated with curvature ones.

Thus, for uncorrelated modes, 6 parameters reduces to $f_{\rm NL}^\zeta$ and 
the combination of $\alpha^2f_{\rm NL}^S$, 
which we investigate in the analysis of this paper.
For correlated modes, we still need 6 parameters.
Although in principle, we can probe all of these parameters/combinations, 
we focus on $f_{\rm NL}^\zeta=\tilde f_{\rm NL}^{\zeta,\zeta\zeta}$ and 
$\alpha f_{\rm NL}^S=\tilde f_{\rm NL}^{S,\zeta\zeta}$ 
for the correlated case since the contributions to the bispectrum from other
isocurvature modes are weak, hence the constraint on $\tilde f_{\rm NL}^{\zeta,\zeta\zeta}$ 
and $\tilde f_{\rm NL}^{S,\zeta\zeta}$ would be most severe. We will again discuss this 
issue after giving a summary of CMB bispectrum in the next subsection.

\subsection{CMB bispectrum}
\label{sec:bispectrum}

Now we briefly summarize the formalism to calculate CMB bispectrum. 
In the analysis, we neglect the secondary non-Gaussianities
arising from the second or higher order cosmological perturbation theory. 

The harmonic coefficients of the CMB temperature anisotropy from 
primordial perturbations $X^A$ can be given as
\begin{equation}
a^A_{lm}=4\pi(-i)^l\int \frac{dk^3}{(2\pi)^3}g^A_l(k)X^A(\vec k)Y^*_{lm}(\hat k), 
\label{eq:alm}
\end{equation}
where $g^A_l (k)$ is the temperature transfer function
for the primordial perturbations $X^A$.
The total CMB anisotropy is the sum of those from the curvature and
isocurvature perturbations, 
i.e. $a_{lm}=a^\zeta_{lm}+a^S_{lm}$.
The angular power spectrum $C_l$, defined by
$\langle a_{lm}{a_{l'm'}}^*\rangle=C_l\delta_{ll'}\delta_{mm'}$,  can be written as
\begin{equation}
C_l=\frac2\pi
\sum_{AB}\int dk\,k^2{g^A_l(k)}{g^B_l(k)}P_{AB}(k).
\end{equation}
It is convenient to define the reduced bispectrum $b_{l_1l_2l_3}$ as
\begin{equation}
\langle a_{l_1m_1}a_{l_2m_2}a_{l_3m_3}\rangle
=b_{l_1l_2l_3}\mathcal G^{l_1l_2l_3}_{m_1m_2m_3}, 
\end{equation}
where $\mathcal G^{l_1l_2l_3}_{m_1m_2m_3}$ is the Gaunt integral, 
which can be written in terms of the Wigner-$3j$ symbol as 
\begin{equation}
\mathcal G^{l_1l_2l_3}_{m_1m_2m_3}=
\sqrt{\frac{(2l_1+1)(2l_2+1)(2l_3+1)}{4\pi}}
\begin{pmatrix}
l_1&l_2&l_3 \\
m_1&m_2&m_3 
\end{pmatrix}
\begin{pmatrix}
l_1&l_2&l_3 \\
0&0&0
\end{pmatrix}.
\end{equation}
Given the bispectrum of primordial perturbations in Eq. \eqref{eq:primb}, 
the reduced CMB bispectrum 
is 
\begin{equation}
b_{l_1l_2l_3}=2\sum_{ABC}f_{\rm NL}^A\int dr r^2\left(
\alpha^A_{l_1}(r)\beta^{B,A}_{l_2}(r)\beta^{C,A}_{l_3}(r)
+(\mbox{2 perms}) 
\right), \label{eq:cmbb}
\end{equation}
where $\alpha^A_l(r)$ and $\beta^{B,A}_l(r)$ are
\begin{eqnarray}
\alpha^A_l(r)&=&\frac2\pi\int dk\,k^2g^A_l(k)j_l(kr), 
\label{eq:alpha} \\
\beta^{B,A}_l(r)&=&\frac2\pi\int dk\,k^2g^B_l(k)j_l(kr)P_{AB}(k).
\label{eq:beta}
\end{eqnarray}
By defining
\begin{equation}
b^{A,BC}_{l_1l_2l_3}=2\int dr r^2\left(
\alpha^A_{l_1}(r)\beta^{B,A}_{l_2}(r)\beta^{C,A}_{l_3}(r)
+(\mbox{2 perms}) 
\right), 
\end{equation}
Eq. \eqref{eq:cmbb} can be recast into
\begin{equation}
b_{l_1l_2l_3}=\sum_{ABC}[f^A_{\rm NL} b^{A,BC}_{l_1l_2l_3}+(\mbox{2 perms})].
\label{eq:btot}
\end{equation}
For later convenience, we introduce a normalized bispectrum
\begin{eqnarray}
\hat b^A_{l_1l_2l_3}
&\equiv& \frac{\partial b_{l_1l_2l_3}}{\partial f^A_{\rm NL}} \\
&=&\sum_{BC}[b^{A,BC}_{l_1l_2l_3}+(\mbox{2 perms})], \notag
\end{eqnarray}
with which we can rewrite Eq. \eqref{eq:btot} as
\begin{equation}
b_{l_1l_2l_3}=\sum_A f_{\rm NL}^A \hat b^A_{l_1l_2l_3}.
\label{eq:btot2}
\end{equation}

When we consider primordial perturbations of uncorrelated isocurvature type, we obtain
\begin{eqnarray}
\hat b^\zeta_{l_1l_2l_3}&=&b^{\zeta,\zeta\zeta}_{l_1l_2l_3}, \\
\hat b^S_{l_1l_2l_3}&=&b^{S,SS}_{l_1l_2l_3}.
\end{eqnarray}
Since  $\hat b^S_{l_1l_2l_3} \propto \alpha^2  f^S_{\rm NL}$, 
 CMB anisotropy is affected by the combination of $\alpha^2 f^S_{\rm NL}$. 
Thus what we can constrain from CMB experiments is $\alpha^2 f^S_{\rm NL}$,  not $f^S_{\rm NL}$ alone.

For the case of correlated isocurvature perturbations, 
we obtain
\begin{eqnarray}
\hat b^\zeta_{l_1l_2l_3}&=&b^{\zeta,\zeta\zeta}_{l_1l_2l_3}
+2b^{\zeta,\zeta S}_{l_1l_2l_3}+b^{\zeta,SS}_{l_1l_2l_3}, 
\label{eq:bz_corr}
\\
\hat b^S_{l_1l_2l_3}&=&b^{S,\zeta\zeta}_{l_1l_2l_3}
+2b^{S,\zeta S}_{l_1l_2l_3}+b^{S,SS}_{l_1l_2l_3}.
\label{eq:bS_corr}
\end{eqnarray}
In terms of $\alpha$, 
the second and third terms in the right hand side in each equation
are respectively of $\mathcal O(\sqrt \alpha)$ and $\mathcal O(\alpha)$ 
higher order than the first terms. From current observations, $\alpha$ 
should be small as  $\alpha\lesssim 0.01$,
hence the first terms in Eqs. \eqref{eq:bz_corr} and \eqref{eq:bS_corr} are leading
in the total bispectrum of Eq. \eqref{eq:btot2}, 
which are proportional to $f_{\rm NL}^\zeta$ and $\alpha f_{\rm NL}^S$, respectively. 
Thus the constraints on these quantities are  most severe among the non-linearity parameters/combinations listed in 
Eq.~\eqref{eq:corr_fNL}.
However, as mentioned in the previous subsection, observations are 
in principle also sensitive to the amplitudes of 
$b^{\zeta,\zeta S}_{l_1l_2l_3},~b^{\zeta, SS}_{l_1l_2l_3},
~b^{S,\zeta S}_{l_1l_2l_3}$ and $b^{S,SS}_{l_1l_2l_3}$, 
which are respectively proportional to $\alpha^{1/2} f_{\rm NL}^\zeta,~\alpha f_{\rm NL}^\zeta,
~\alpha^{3/2} f_{\rm NL}^S$
and $ \alpha^{2} f_{\rm NL}^S$. 
Although their contribution could also affect constraints on $f_{\rm NL}$ and 
$\alpha f_{\rm NL}^{\rm (ISO)}$, nevertheless, such an effect is expected to be 
weak as far as $\alpha$ is small as
$\alpha\lesssim0.01$.  We will come back to this issue at the end of Section \ref{sec:result}.

\subsection{Fisher matrix analysis}
\label{sec:fisher}

Before presenting our constraints from observed data, 
we here present expected errors from WMAP7 data
based on the Fisher matrix analysis, which serves as a cross-check of our results 
given in Section \ref{sec:result}.
 
  According to Refs. 
\cite{Komatsu:2001rj,Babich:2004yc}, 
given the bispectrum of Eq. \eqref{eq:btot2}, 
the Fisher matrix for the 
bispectrum  is
given by
\begin{equation}
F_{AB}=\frac{f_{\rm sky}}6
\sum_{l_{\rm max}\ge l_1,l_2,l_3\ge2}\frac{(2l_1+1)(2l_2+1)(2l_3+1)}{4\pi}
\begin{pmatrix}
l_1 & l_2 & l_3 \\
0 & 0 & 0 &
\end{pmatrix}^2
\frac{\hat b^A_{l_1l_2l_3}\hat b^B_{l_1l_2l_3}}{\mathcal C_{l_1}\mathcal C_{l_2}\mathcal C_{l_3}}, 
\end{equation}
where $\mathcal C_l=C_l+N_l$ with $C_l$ and $N_l$ being 
 the power spectra of signal and noise, respectively.
Given a survey with a finite resolution, $F_{AB}$ saturates as $l_{\rm max}$ 
are increased.

The signal power spectrum $C_l$ are computed using the CAMB code \cite{Lewis:1999bs}
assuming the mean parameters from the WMAP 7-year data alone \cite{Komatsu:2010fb}.
We set the same spectral indices for both the power spectra of
adiabatic and isocurvature perturbations.
On the other hand, the noise power spectrum can be approximated with
 the Knox's formula \cite{Knox:1995dq} as,
\begin{equation}
N_l=\theta_{\rm FWHM}^2\sigma_T^2\exp\left[l(l+1)\frac{\theta_{\rm FWHM}^2}{8\ln2}\right],
\end{equation}
where $\theta_{\rm FWHM}$ is the 
full width at half maximum of the Gaussian beam, and $\sigma_T$ is the
root mean square of the instrumental noise par pixel.
The total $N_l$ of a multi-frequency survey can be given by 
a quadrature sum of $N_l$ of each frequency band.
For the Fisher matrix analysis, we adopt the specification of 
 the 7-year  WMAP V and W bands for $\theta_{\rm FWHM}$ and $\sigma_T$,
which are are listed in Table \ref{tbl:wmap7} \cite{wmap}.
Furthermore, we assume that $f_{\rm sky}=0.72$, 
which is consistent with the mask we adopt in Section \ref{sec:result}.

\begin{table}[htb]
  \begin{center}
  \begin{tabular}{llcc}
  \hline
  \hline
  band & V & W \\
  \hline
  $\theta_{\rm FWHM}$ [arcmin] & 21.0 & 13.2 \\
  $\sigma_T$ [mK] & 25 & 38 \\
  \hline
  \hline 
\end{tabular}
  \caption{
  Survey parameters for 7-year observation of the WMAP V and W bands.
  }
  \label{tbl:wmap7}
\end{center}
\end{table}

From $F_{AB}$, we can evaluate ideal constraints on
the nonlinearity parameters. For a single nonlinearity parameter, 
two different constraints can be derived. One is the non-marginalized 
constraint in which we assume that the other nonlinearity parameter
is  fixed to zero. In this case, 1 $\sigma$ error 
should be given by $\Delta f^A_{\rm NL}=1/\sqrt{F_{AA}}$.
Another one is the marginalized constraint
which assumes the other nonlinearity parameter varies freely.
In this case, 1$\sigma$ error is given by $\Delta f^A_{\rm NL}=\sqrt{F^{-1}_{AA}}$.
In Table \ref{tbl:fisher_uncorr}, we list ideal 1$\sigma$ constraints 
from the WMAP 7-year observation
for uncorrelated CI and NID models.
Those for the correlated CI and NID models are listed in Table \ref{tbl:fisher_corr}.
These results are to be consulted for validity check of 
our constraints from actual data in Section \ref{sec:result}. 

\begin{table}[htb]
  \begin{center}
  \begin{tabular}{llcc}
  \hline
  \hline
  & $\Delta f_{\rm NL}$ & $\Delta (\alpha^2 f_{\rm NL}^{\rm (ISO)})$ \\
  \hline
  uncorrelated CI & 20 (22) & 65 (72) \\
  uncorrelated NID & 20 (38) & 135 (256) \\ 
  \hline
  \hline 
\end{tabular}
  \caption{
  1$\sigma$ errors evaluated from the Fisher matrix analysis on the nonlinearity parameters 
  from the WMAP 7-year observation for the uncorrelated CI and NID models.
  Errors with and without parenthesis correspond to non-marginalized
  and marginalized constraints, respectively. }
  \label{tbl:fisher_uncorr}
\end{center}
\end{table}
\begin{table}[htb]
  \begin{center}
  \begin{tabular}{llcc}
  \hline
  \hline
  & $\Delta f_{\rm NL}$ & $\Delta (\alpha f_{\rm NL}^{\rm (ISO)})$ \\
  \hline
  correlated CI & 20 (24) & 120 (140) \\
  correlated NID & 20 (56) & 51 (143) \\ 
  \hline
  \hline 
\end{tabular}
  \caption{
  The same as in Table \ref{tbl:fisher_uncorr} except 
  for the correlated CI and NID models.
  }
  \label{tbl:fisher_corr}
\end{center}
\end{table}

\section{Analysis method} 
\label{sec:analysis}

Our analysis method is the same as in our previous paper 
\cite{Hikage:2012be}. 
For more details, we refer to Ref. \cite{Hikage:2012be} and 
references therein.

We adopt the following optimal estimator of nonlinearity parameters $f_{\rm NL}^A$, 
\begin{equation}
\begin{pmatrix}
\hat f^\zeta_{\rm NL} \\
\hat f^S_{\rm NL}
\end{pmatrix}=
\begin{pmatrix}
\langle S_{\rm prim}^\zeta\rangle_{f^\zeta_{\rm NL}=1} &
\langle S_{\rm prim}^\zeta\rangle_{f^S_{\rm NL}=1} \\
\langle S_{\rm prim}^S\rangle_{f^\zeta_{\rm NL}=1} & 
\langle S_{\rm prim}^S\rangle_{f^S_{\rm NL}=1}
\end{pmatrix}^{-1}
\begin{pmatrix}
S_{\rm prim}^\zeta \\
S_{\rm prim}^S
\end{pmatrix},
\label{eq:estim2}
\end{equation}
where a square bracket with  subscript $f^A_{\rm NL}=1$ indicates 
an ensemble average over simulations with
non-zero non-linearity parameter shown in the subscript and
the other non-linearity parameter being fixed to zero.
$S^A_{\rm prim}$ is the cubic statistics
\cite{Komatsu:2003iq,Yadav:2007rk,
Creminelli:2005hu,Yadav:2007ny}, which is given by
\begin{equation}
S^A_{\rm prim}=\frac16\sum_{\{lm\}}
\hat b^A_{l_1l_2l_3}
\mathcal G^{l_1l_2l_3}_{m_1m_2m_3}
\left[\tilde a_{l_1 m_1}
\tilde a_{l_2 m_2}
\tilde a_{l_3 m_3}-3\tilde a_{l_1 m_1}
\langle\tilde a_{l_2 m_2}
\tilde a_{l_3 m_3}\rangle_0
\right], 
\label{eq:cubic}
\end{equation}
where the square bracket with subscript $0$ indicates an ensemble average 
over Gaussian simulations.
$\tilde a_{lm}$ is a harmonic coefficient of observed (or simulated) data
maps weighted by inverse-variance,
\begin{equation}
\tilde a_{lm}\equiv\sum_{l'm'}\mathcal C_{lm,l'm'}^{-1}a_{l'm'}, 
\end{equation}
which is derived from the optimality of the estimator of Eq.~\eqref{eq:estim2} \cite{Smith:2009jr}. 
Eq.~\eqref{eq:estim2} can be schematically represented as 
$\hat f^A_{\rm NL}=\displaystyle\sum_B
\langle S^A_{\rm prim}\rangle_{f^{B}_{\rm NL}=1}^{-1}
S^B_{\rm prim}$.
In our analysis, to evaluate $\langle \tilde a_{lm}\tilde a_{l'm'}\rangle_0$, 
we accumulated at least $250$ Monte Carlo (MC) samples.

Assuming the Gaussianity for $X^A$, we estimate 
the covariance of the estimator $\hat f^A_{\rm NL}$, 
\begin{eqnarray}
\langle \hat f^A_{\rm NL}\hat f^B_{\rm NL}\rangle_0&=&\sum_{A'B'}
\langle S^A_{\rm prim}\rangle_{f^{A'}_{\rm NL}=1}^{-1}
\langle S^{A'}_{\rm prim}S^{B'}_{\rm prim}\rangle_0
\langle S^B_{\rm prim}\rangle_{f^{B'}_{\rm NL}=1}^{-1} \\
&=&\langle S^A_{\rm prim}\rangle_{f^{B}_{\rm NL}=1}^{-1}. \notag
\end{eqnarray}
Here we used the relation 
$\langle S^A_{\rm prim}\rangle_{f^B_{\rm NL}=1}=
\langle S^A_{\rm prim}S^B_{\rm prim}\rangle_0$, which has been proved  
in our previous paper \cite{Hikage:2012be}.
When we assume that either of adiabatic or uncorrelated isocurvature perturbations
are non-Gaussian and the other is Gaussian, Eq.~\eqref{eq:estim2}
can be reduced to
\begin{equation}
\hat f^A_{\rm NL}=S_{\rm prim}^A/\langle S_{\rm prim}^A\rangle_{f^A_{\rm NL}=1},
\end{equation}
where the subscripts $A$ indicates the perturbations which are assumed to be 
non-Gaussian. In this case, the variance of $\hat f^A_{\rm NL}$ is given by 
$1/\langle S^A_{\rm prim}\rangle_{f^A_{\rm NL}=1}$.

In order to determine the normalization $\langle S^A_{\rm prim}\rangle_{f^B_{\rm NL}=1}$, 
we need to simulate non-Gaussian CMB maps.
We adopt the method for local-type non-Gaussianity 
developed in Ref. \cite{Elsner:2009md}, in which  
the quadrature method with optimized nodes and weights 
for line of sight integral is used. 
In our analysis, we require the method to be accurate such that
the mean square of the error is less than 0.01 at each multipole $(lm)$.
For $l_{\rm max}=1024$, we found that this level of accuracy requires 42, 
15 and 14 nodes for adiabatic, CI and NID perturbations, respectively.

Computation of inverse-variance weighted map $\tilde a_{lm}$ 
is performed based on the method of Ref. \cite{Smith:2007rg}, 
which adopts a conjugate gradient method with multi-grid
preconditioning. This method also allows us to marginalize over 
the amplitude of components which are proportional to 
template maps. 

\section{Result}
\label{sec:result}

In our analysis, we assume a flat power-law $\Lambda$CDM model as a
fiducial one and adopt the mean values for cosmological parameters from the 
WMAP 7-year data alone \cite{Komatsu:2010fb} 
\begin{equation}
(\Omega_b,\Omega_c, H_0, \tau, A_{\rm adi})=
(0.0448, 0.220, 71, 0.088, 2.43\times 10^{-9}).
\end{equation}
Here $\Omega_b$ and $\Omega_c$ are the density parameters for 
baryon and CDM, respectively. $H_0$ is the Hubble constant 
in units of  km/sec/Mpc and $\tau$ is the optical depth for reionization.
$A_{\rm adi}$ is the amplitude of power spectrum of 
adiabatic perturbations at the reference scale $k_*=0.002$Mpc$^{-1}$. 
With this definition, the power spectrum can be written as
$\mathcal{P}_{\zeta\zeta} (k)= (k^3/2\pi^2) P_{\zeta\zeta}(k)=A_{\rm adi}(k/k_*)^{n_{\rm adi}-1}$, where
$n_{\rm adi}$ is the spectral index for $P_{\zeta\zeta}$.
We assume that the power spectrum of isocurvature perturbations
is also given in a power-law form $\mathcal{P}_{S_{\rm G}S_{\rm G}} (k)
= (k^3/2\pi^2) P_{S_{\rm G}S_{\rm G}}(k)\propto k^{n_{\rm iso}-1}$.
With regard to the values of $n_{\rm adi}$ and $n_{\rm iso}$, we adopt several different 
settings, which will be separately presented when we show our constraints.
The transfer function of CMB is computed using 
the CAMB code \cite{Lewis:1999bs}.
We combine the foreground-cleaned maps of V and W bands of the WMAP 7-year data 
\cite{Jarosik:2010iu,Gold:2010fm}\footnote{http://lambda.gsfc.nasa.gov}
with a resolution $N_{\rm side}=512$ in the HEALPix pixelization scheme \cite{Gorski:2004by}\footnote{http://healpix.jpl.nasa.gov}. We adopt the KQ75y7 mask \cite{Gold:2010fm}, 
which cuts 28.4 \% of the sky. We also set the maximum multipole 
$l_{\rm max}$ to 1024 in our analysis. 
We marginalize the amplitudes of the monopole $l=0$ and dipoles $l=1$ as default
and also optionally marginalize the amplitudes of the Galactic foreground components at large angular
scales using the synchrotron, free-free and dust emission templates from Ref. \cite{Gold:2010fm}.

\begin{figure}
  \begin{center}
    \begin{tabular}{cc}
      \hspace{0mm}\scalebox{.9}{\includegraphics{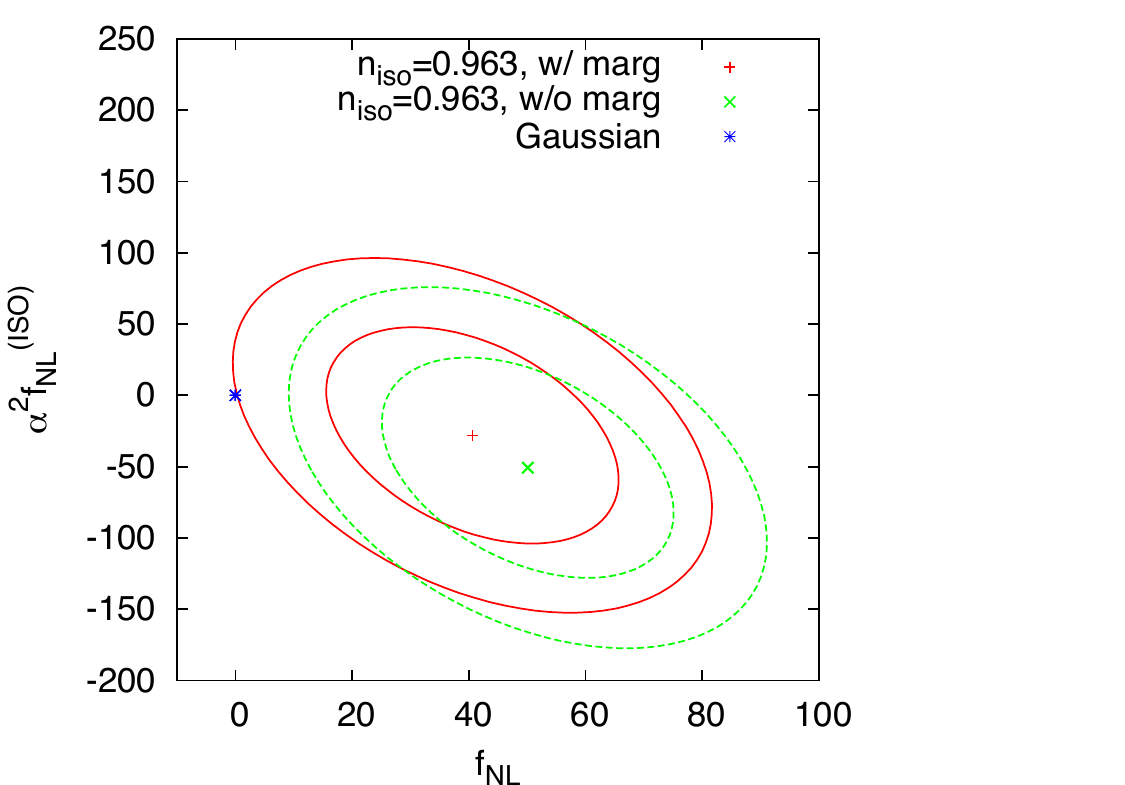}} &
      \hspace{-30mm}\scalebox{.9}{\includegraphics{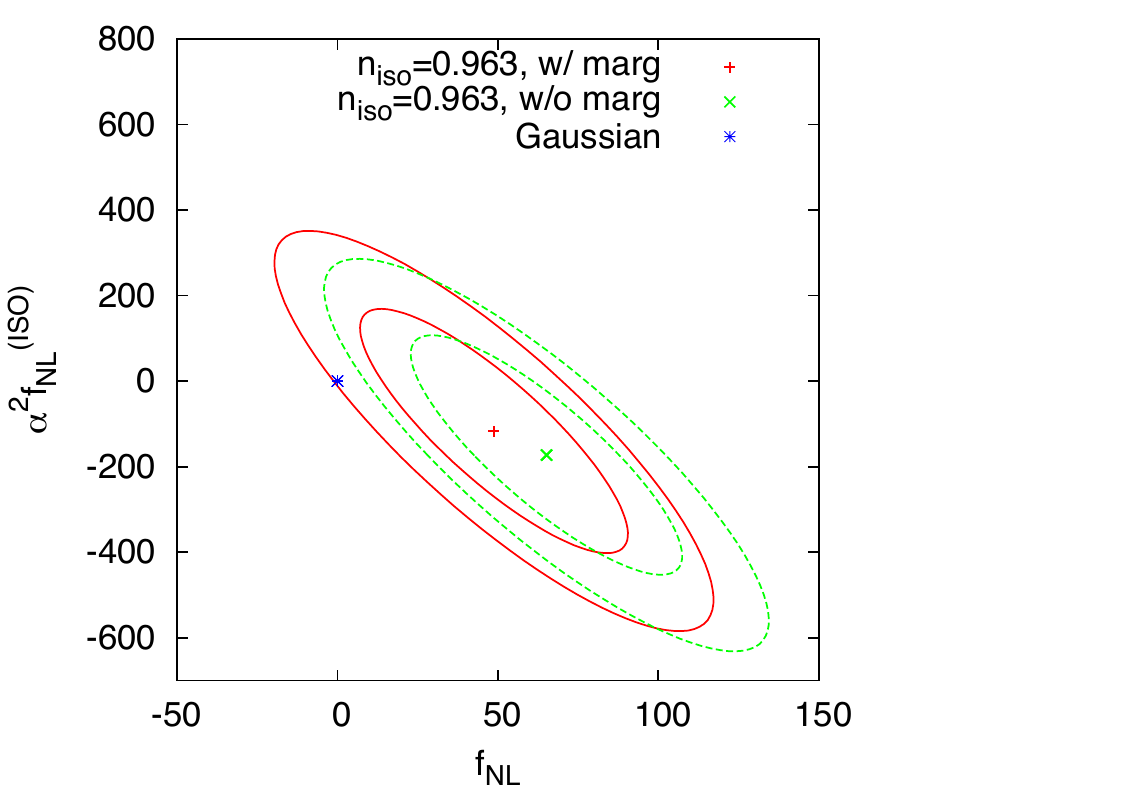}} \\
      \hspace{0mm}\scalebox{.9}{\includegraphics{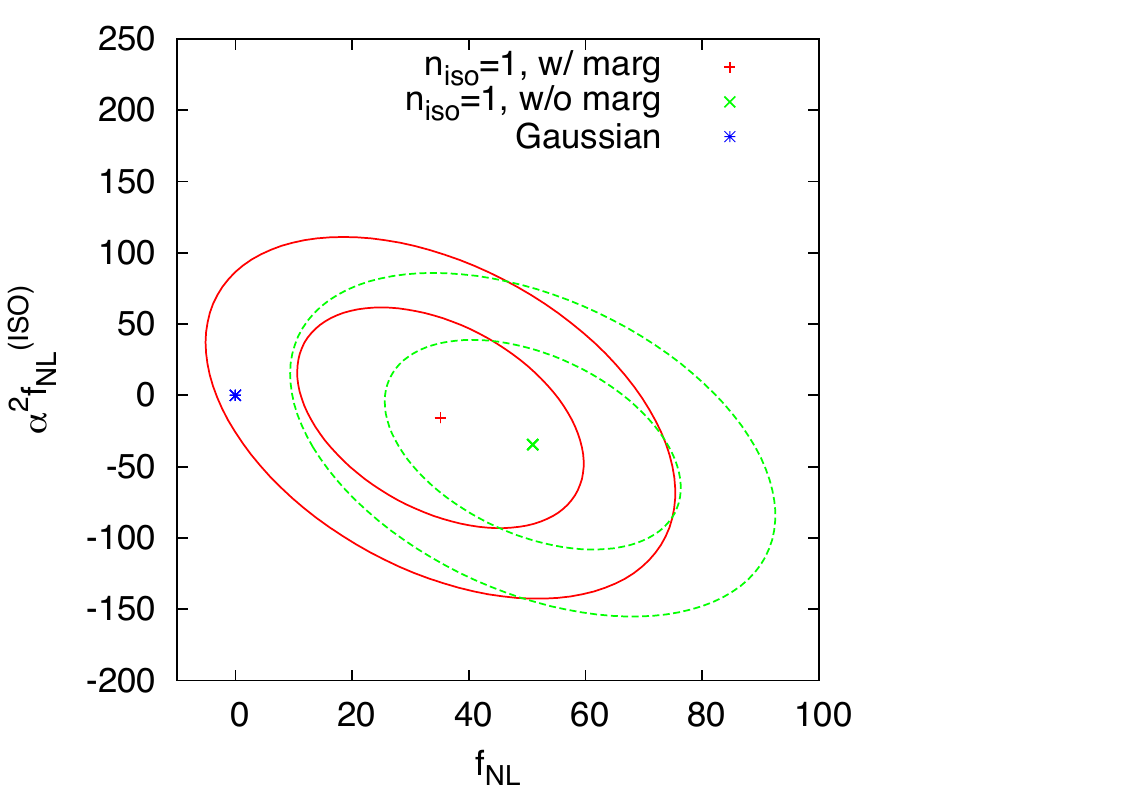}} &
      \hspace{-30mm}\scalebox{.9}{\includegraphics{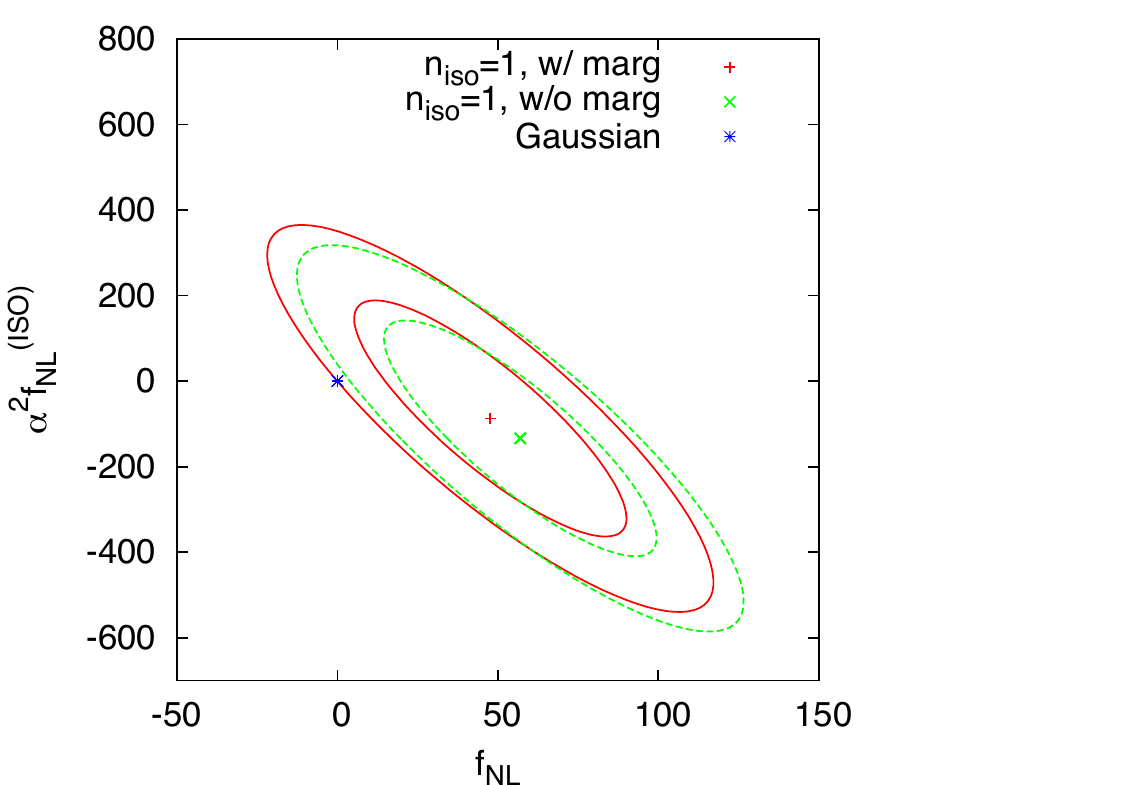}}
    \end{tabular}
  \end{center}
  \caption{
  Joint constraints on $f_{\rm NL}$ and 
  $\alpha^2f_{\rm NL}^{\rm (ISO)}$ 
  at 1 and 2$\sigma$ levels for uncorrelated isocurvature models. 
  Different panels show cases for different isocurvature modes and 
  values of $n_{\rm iso}$.
  Left and right panels respectively correspond to the cases of CI and NID modes.
  Top and bottom panels respectively correspond to the cases of $n_{\rm iso}=0.963$
  and $n_{\rm iso}=1$. Here $n_{\rm adi}$ is fixed to $0.963$.
  In each panel, solid red (green dashed) lines are the constraints with (without)
  template marginalization of foregrounds.
  Blue star corresponds to the case where primordial perturbations are purely Gaussian.
  }
  \label{fig:uncorr}
\end{figure}

Constraints for the case of uncorrelated isocurvature perturbations
are listed in Table~\ref{tbl:uncorr}. Concerning the spectral indices, 
we fix $n_{\rm adi}$ to the WMAP 7-year mean value $0.963$, while
for $n_{\rm iso}$, we adopt two settings, $n_{\rm iso}=0.963$ and 
$n_{\rm iso}=1$, which are separately analyzed.
Joint constraints on $f_{\rm NL}=\frac53f_{\rm NL}^\zeta$
and $\alpha^2f^{\rm (ISO)}_{\rm NL}=\alpha^2f^S_{\rm NL}$ 
are shown in Fig.~\ref{fig:uncorr}. 
First we found that the 1$\sigma$ errors are roughly the same 
as those obtained from the Fisher matrix analysis given in Table~\ref{tbl:fisher_uncorr}
regardless of marginalization of foregrounds and fiducial $n_{\rm iso}$.
The fiducial values of $n_{\rm iso}$ little affect the central values for both the
uncorrelated CI and NID models. 
While the marginalization of foregrounds can bias the central values in
the uncorrelated CI model by as much as 0.5$\sigma$, 
the effects in the uncorrelated NID model are smaller, 
from which we can see that 
the effects of residual foregrounds are moderate.
To conclude the case for the uncorrelated modes, the primordial perturbations are consistent with Gaussian
ones at around 2$\sigma$ level, even if we consider uncorrelated 
CI or NID perturbations.

\begin{table}[htb]
  \begin{center}
  \begin{tabular}{llcc}
  \hline
  \hline
  \multicolumn{2}{c}{setups}& $f_{\rm NL}$ & $\alpha^2 f_{\rm NL}^{\rm (ISO)}$ \\
  \hline
  CI, $n_{\rm iso}=0.963$
  &w/o template marginalization 
  & $43\pm21$ &$13\pm66$ \\
  && $(50\pm23)$ & $(-51\pm 72)$\\
  \cline{2-4} 
  &w/ template marginalization 
  & $37\pm21$ & $22\pm64$ \\
  && $(41\pm23)$ & $(-28\pm71)$ \\
  \hline
  CI, $n_{\rm iso}=1$ 
  &w/o template marginalization 
  & $46\pm21$ & $26\pm63$ \\
  && $(51\pm23)$ &  $(-34\pm69)$ \\
  \cline{2-4} 
  &w/ template marginalization 
  & $33\pm21$ & $30\pm66$\\
  && $(35\pm23)$ & $(-15\pm72)$ \\
  \hline
  NID,  $n_{\rm iso}=0.963$
  &w/o template marginalization 
  & $43\pm21$ &$191\pm140$ \\
  && $(65\pm39)$ & $(-173\pm 261)$\\
  \cline{2-4} 
  &w/ template marginalization 
  & $34\pm21$ & $164\pm143$ \\
  && $(48\pm39)$ & $(-116\pm266)$ \\
  \hline
  NID,  $n_{\rm iso}=1$
  &w/o template marginalization 
  & $40\pm21$ &$178\pm137$ \\
  && $(57\pm40)$ & $(-133\pm 257)$\\
  \cline{2-4} 
  &w/ template marginalization 
  & $36\pm21$ & $175\pm137$ \\
  && $(48\pm40)$ & $(-87\pm257)$ \\
  \hline
  \hline 
\end{tabular}
  \caption{
  Constraints on $f_{\rm NL}$ and $\alpha^2 f_{\rm NL}^{\rm (ISO)}$ at 1$\sigma$ level
  for the cases of uncorrelated isocurvature perturbations.
  A value with (without) parenthesis is a constraint on a nonlinearity parameter
  without (with) marginalization of the other one. 
  }
  \label{tbl:uncorr}
\end{center}
\end{table}

\begin{figure}
  \begin{center}
    \begin{tabular}{cc}
      \hspace{0mm}\scalebox{.9}{\includegraphics{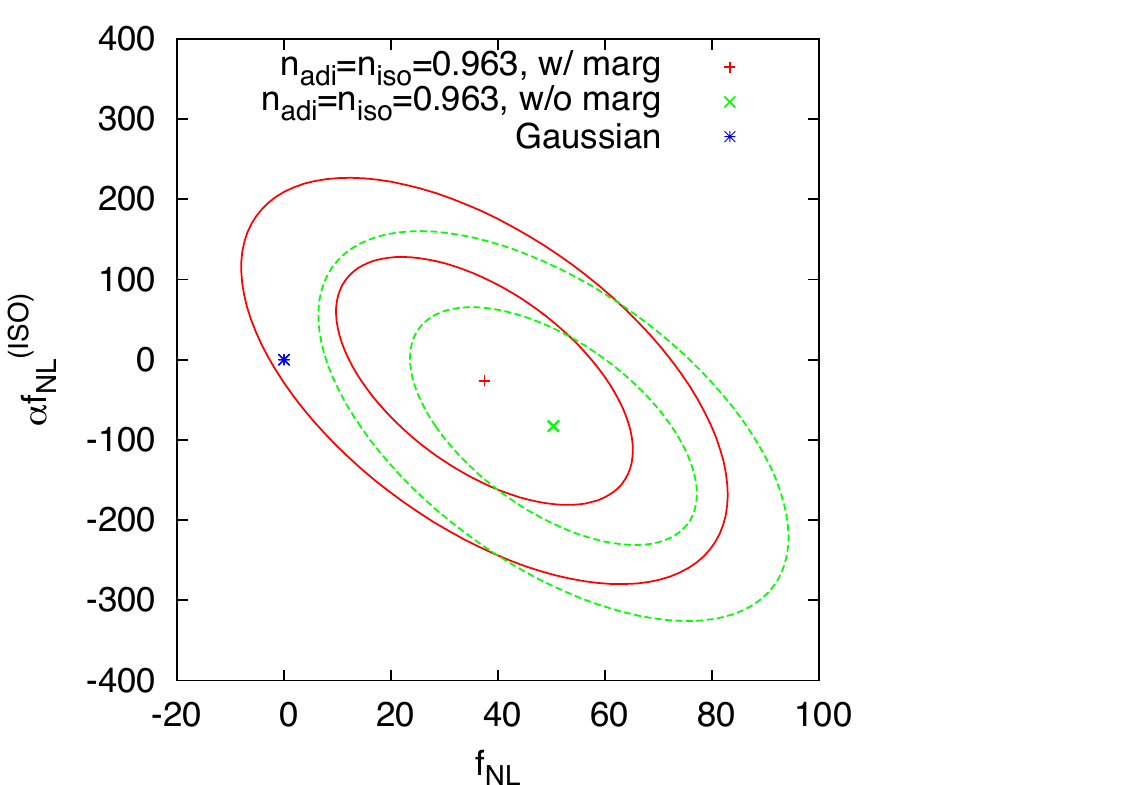}} &
      \hspace{-30mm}\scalebox{.9}{\includegraphics{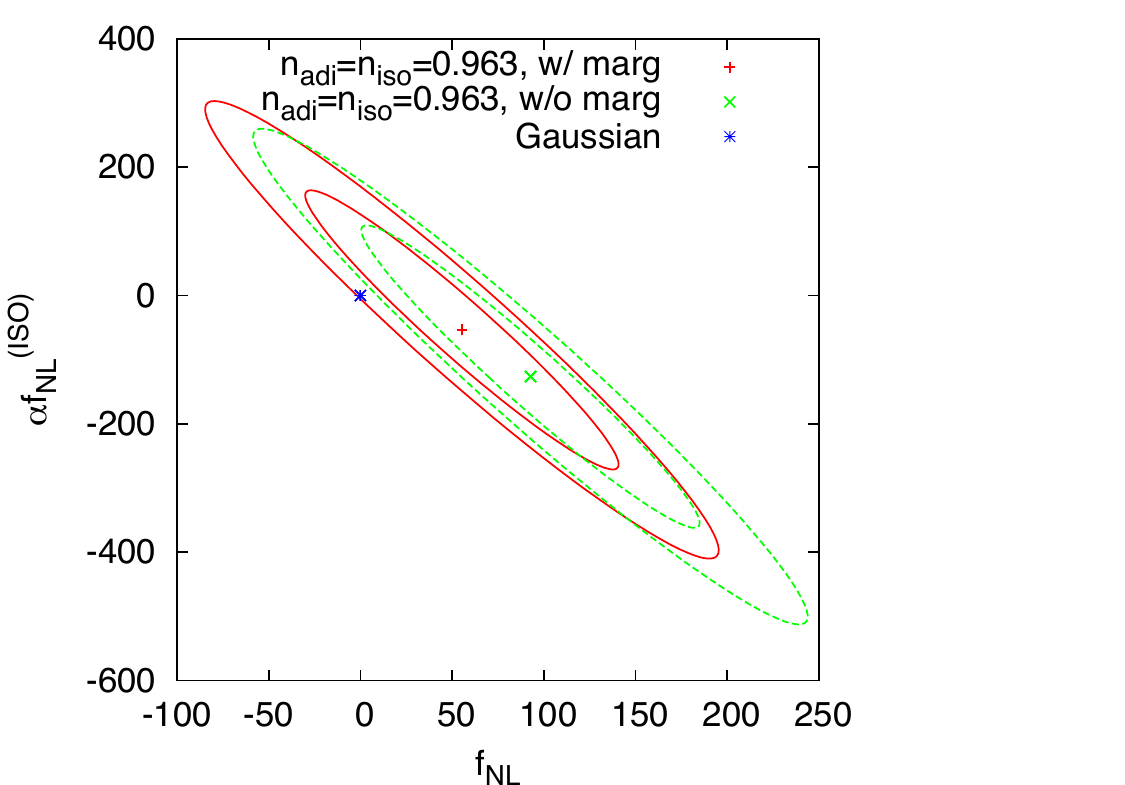}} \\
      \hspace{0mm}\scalebox{.9}{\includegraphics{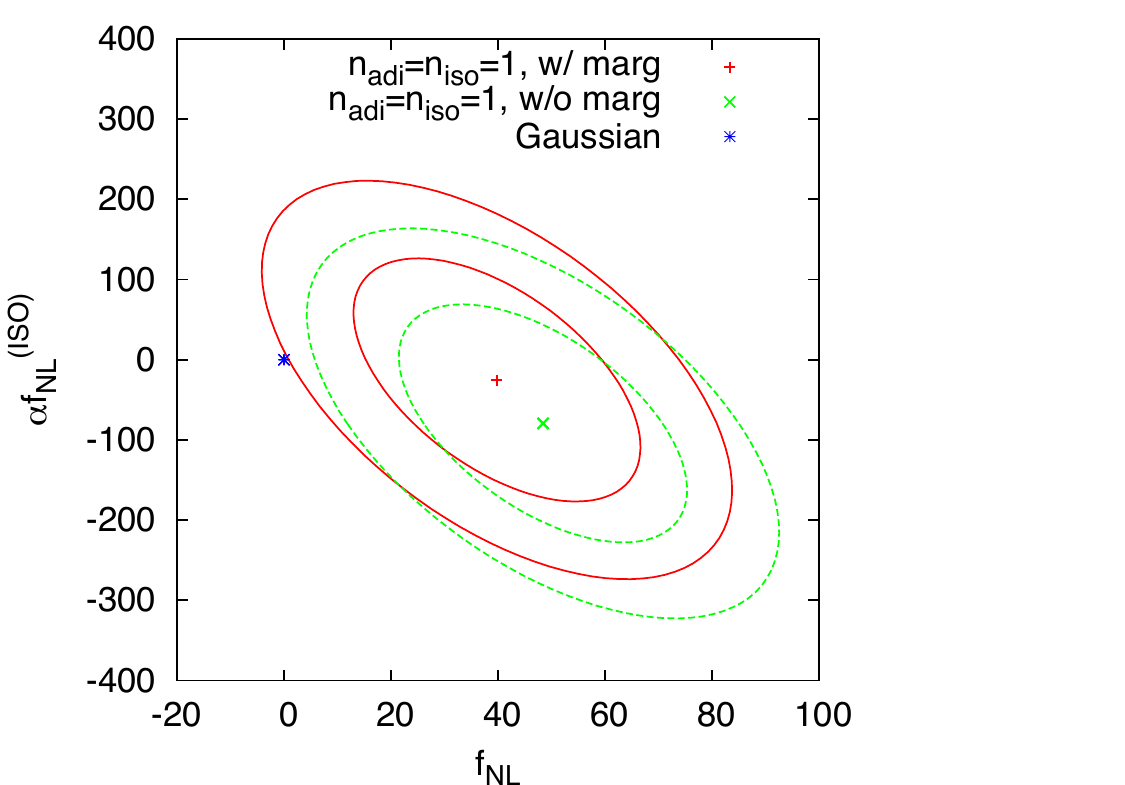}} &
      \hspace{-30mm}\scalebox{.9}{\includegraphics{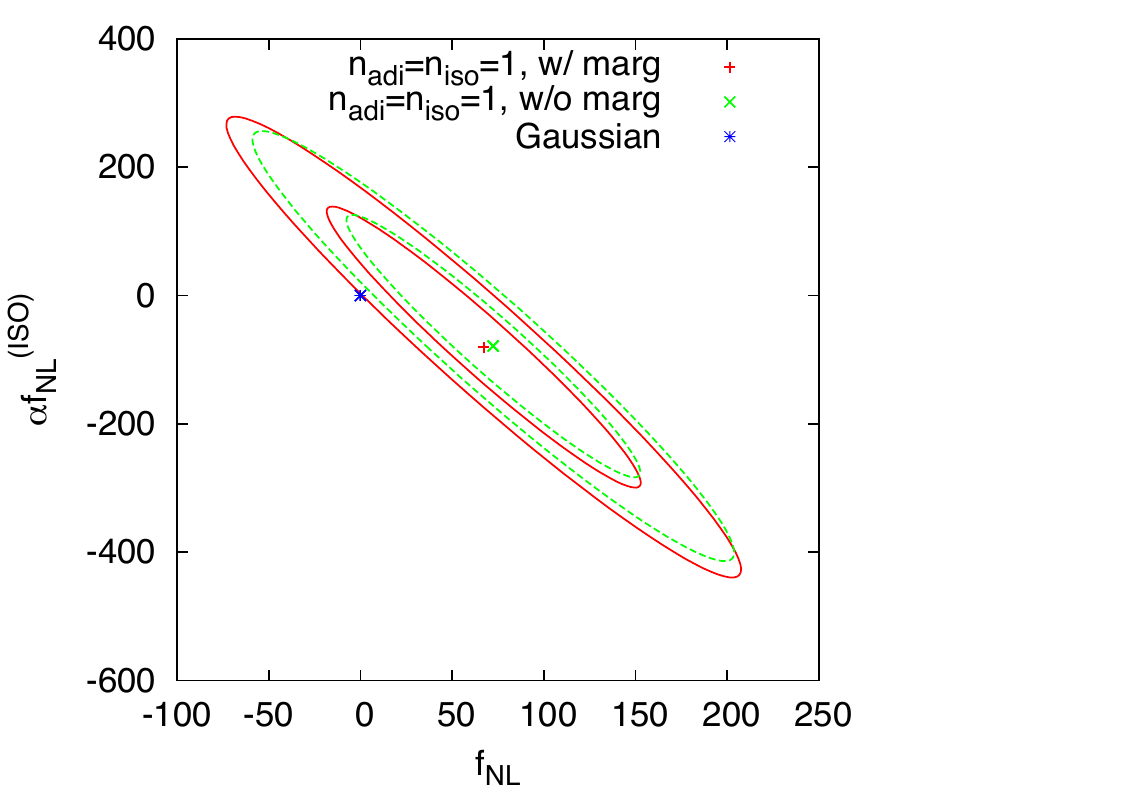}}
    \end{tabular}
  \end{center}
  \caption{
  Joint constraints on $f_{\rm NL}$ and 
  $\alpha f_{\rm NL}^{\rm (ISO)}$ 
  at 1 and 2$\sigma$ levels for correlated isocurvature models. 
  Left and right panels show the constraints for the cases of CI and NID modes.
  We here assume $n_{\rm adi}=n_{\rm iso}$ and cases of $n_{\rm adi}=n_{\rm iso}=0.963$
  and $1$ are respectively shown in top and bottom panels. 
  Other conventions are the same as in Fig. \ref{fig:uncorr}.
  }
  \label{fig:corr}
\end{figure}

Constraints for the case of correlated isocurvature perturbations
are listed in Table~\ref{tbl:uncorr}. 
With regard to the values of spectral indices, we adopt two different
cases, $n_{\rm adi}=n_{\rm iso}=0.963$ and $n_{\rm adi}=n_{\rm iso}=1$ 
for correlated case.
Joint constraints on $f_{\rm NL}=\frac53f_{\rm NL}^\zeta$ and $\alpha f^{\rm (ISO)}_{\rm NL}
=\alpha f_{\rm NL}^S$ are 
shown in Fig.~\ref{fig:corr}. We found that the resulting constraints  
are consistent with the Fisher matrix analysis shown 
in Table~\ref{tbl:fisher_corr} in this case as well, although marginalized errors
for the NID model are substantially larger than the Fisher analysis, 
which may result from the inhomogeneous noise and the sky cut.
We found that fiducial values of the spectral indices little
affect the constraints. On the other hand, template marginalization 
can bias the central values by as much as 0.5$\sigma$,
in particular for CI model. However, as in the uncorrelated case, 
the effects of residual foregrounds are regarded not to be severe at all. We can conclude that 
the primordial perturbations
are consistent with Gaussian ones at around 2$\sigma$ level also 
 in the case of correlated CI and NID models. 

Finally we comment on the contributions from the terms which are higher order with
$\mathcal{O} (\sqrt{\alpha})$ and $\mathcal{O}  (\alpha )$ 
such as the second and third terms in Eqs.~\eqref{eq:bz_corr} and \eqref{eq:bS_corr}.
In the analysis of correlated isocurvature models, 
we have neglected those terms since they are expected to give little effects on the constraints. 
To check this, we have also made an analysis 
where we included those terms.   
We have compared the constraints with and without those terms for the case of $\alpha=0.003$. 
It is found that the errors are almost unchanged and
the central values change by no more than  $\mathcal{O}(0.1)\sigma$,
which justifies ignoring higher order terms in $\alpha$ in our analysis.

\begin{table}[htb]
  \begin{center}
  \begin{tabular}{llcc}
  \hline
  \hline
  \multicolumn{2}{c}{setups}& $f_{\rm NL}$ & $\alpha f_{\rm NL}^{\rm (ISO)}$ \\
  \hline
  CI,  $n_{\rm iso}=n_{\rm adi}=0.963$
  &w/o template marginalization 
  & $41\pm21$ &$76\pm114$ \\
  && $(50\pm25)$ & $(-82\pm 138)$\\
  \cline{2-4} 
  &w/ template marginalization 
  & $34\pm21$ & $90\pm120$ \\
  && $(37\pm25)$ & $(-26\pm144)$ \\
  \hline
  CI,  $n_{\rm iso}=n_{\rm adi}=1$
  &w/o template marginalization 
  & $40\pm21$ &$70\pm114$ \\
  && $(48\pm25)$ & $(-79\pm 138)$\\
  \cline{2-4} 
  &w/ template marginalization 
  & $37\pm21$ & $99\pm117$ \\
  && $(40\pm25)$ & $(-25\pm141)$ \\
  \hline
  NID,  $n_{\rm iso}=n_{\rm adi}=0.963$
  &w/o template marginalization 
  & $45\pm21$ &$103\pm55$ \\
  && $(93\pm86)$ & $(-126\pm 220)$\\
  \cline{2-4} 
  &w/ template marginalization 
  & $35\pm21$ & $82\pm54$ \\
  && $(55\pm80)$ & $(-53\pm203)$ \\
  \hline
  NID,  $n_{\rm iso}=n_{\rm adi}=1$
  &w/o template marginalization 
  & $42\pm21$ &$99\pm53$ \\
  && $(72\pm75)$ & $(-78\pm 191)$\\
  \cline{2-4} 
  &w/ template marginalization 
  & $36\pm21$ & $86\pm53$ \\
  && $(67\pm80)$ & $(-80\pm204)$ \\
  \hline
  \hline 
\end{tabular}
  \caption{
  Constraints on $f_{\rm NL}$ and $\alpha f_{\rm NL}^{\rm (ISO)}$ 
  for the cases of correlated isocurvature perturbations.
  }
  \label{tbl:corr}
\end{center}
\end{table}

\section{Implications to the curvaton scenario} \label{sec:curvaton}
In this section, we demonstrate an application of our result to a model of the very early Universe 
based on the curvaton scenario.
We here focus on a  model
where inflaton and curvaton fields source the primordial curvature
and CDM isocurvature perturbations.
In addition, we assume that some fraction of CDM is 
generated before curvaton decay, and the rest of CDM
is generated directly from the curvaton decay. No CDM production after
curvaton decay is assumed.
Here we briefly outline the model and its predictions
for the power spectrum and non-linearity parameters. 
For more details, we refer to e.g. Refs. 
\cite{Langlois:2008vk,Kawasaki:2008pa,Langlois:2011zz,Langlois:2010fe}.

Adopting the sudden decay approximation, 
the curvaton decays synchronously on the uniform-density 
hypersurface of the total matter. On this hypersurface, 
the energy conservation of the total matter and CDM
should be given by
\begin{eqnarray}
\bar\rho_r&=& 
\bar\rho^{(\phi)}_r e^{4(\zeta^{(\phi)} -\zeta)}+\bar\rho_\sigma e^{3(\zeta_\sigma-\zeta)}, 
\label{eq:rad}
\\
\bar\rho_{\rm CDM} e^{3(\zeta_{\rm CDM}-\zeta)}&=&
\bar\rho^{(\phi)}_{\rm CDM} e^{3(\zeta^{(\phi)}-\zeta)}
+\bar\rho^{(\sigma)}_{\rm CDM} e^{3(\zeta_\sigma-\zeta)},
\label{eq:mat}
\end{eqnarray}
where $\rho_r^{(\phi)}$ and 
$\rho_\sigma$ are respectively energy densities of radiation and curvaton 
just before decay. 
The subscript $\phi$ for $\rho_r^{(\phi)}$ is put to emphasize that the radiation 
originates from the inflaton decay (or decay products).
$\rho_r$ and $\rho_{\rm CDM}$ are respectively ones 
for radiation and CDM after the decay.
$\rho_{\rm CDM}^{(\phi)}$ and $\rho_{\rm CDM}^{(\sigma)}$ are
respectively energy densities  of CDM produced from the inflaton and the curvaton, 
 which leads to $\bar\rho_{\rm CDM}=
\bar\rho_{\rm CDM}^{(\phi)}+\bar\rho_{\rm CDM}^{(\sigma)}$.
$\zeta^{(\phi)}$ is the curvature perturbation generated from the inflaton.
$\zeta_\sigma$ and $\zeta_{\rm CDM}$ respectively are the ones on uniform curvaton and CDM 
energy density hypersurfaces. $\zeta$ is for the total one.

In the first equation, we assumed that  CDM 
is energetically negligible during the decay and also  the Universe is almost completely
dominated by radiation after the curvaton decay. Otherwise, the Universe is dominated by
matter soon after its decay, which would require further entropy production.

To describe the model, we introduce some parameters. 
We define the fractions of energy densities created from the curvaton decay 
in radiation and CDM as $\epsilon_r$ and $\epsilon_{\rm CDM}$, respectively, i.e.
\begin{equation}
\epsilon_r\equiv\bar\rho_\sigma/\bar\rho_r,~
\qquad\qquad
\epsilon_{\rm CDM} \equiv\bar\rho^{(\sigma)}_{\rm CDM}/\bar\rho_{\rm CDM}.
\end{equation}
Then Eqs.~\eqref{eq:rad} and \eqref{eq:mat}
can be recast into
\begin{eqnarray}
1&=&(1-\epsilon_r)e^{4(\zeta^{(\phi)}-\zeta)}+\epsilon_r e^{3(\zeta_\sigma-\zeta)}, 
\label{eq:rad2}\\
e^{3(\zeta_{\rm CDM}-\zeta)}&=&
(1-\epsilon_{\rm CDM})e^{3(\zeta^{(\phi)}-\zeta)}
+\epsilon_{\rm CDM} e^{3(\zeta_\sigma-\zeta)}.
\label{eq:mat2}
\end{eqnarray}

By solving Eqs. \eqref{eq:rad2} and \eqref{eq:mat2} 
for the curvature perturbation $\zeta$ and CDM isocurvature fluctuations 
$S_{\rm CDM}=3(\zeta_{\rm CDM}-\zeta)$
up to second order in $\zeta^{(\phi)}$ and $S_\sigma = 3 (\zeta_\sigma - \zeta^{(\phi)})$, 
we obtain 
\begin{eqnarray}
\zeta&=&\zeta^{(\phi)}+\frac{rS_\sigma}3+
\frac{(1-r)(3+r)}{2r} \left(\frac{rS_\sigma}3 \right)^2, 
\label{eq:zeta_cur}\\
S_{\rm CDM}&=&
\left[\frac{\epsilon_{\rm CDM}}{r}-1\right]rS_\sigma\notag \\
&&+\left[
\frac{\epsilon_{\rm CDM}(1-\epsilon_{\rm CDM})}{2r^2}
-\frac{(1-r)(3+r)}{6r}
\right]\left(rS_\sigma\right)^2,
\label{eq:S_cur}
\end{eqnarray}
where $r$ is defined as
\begin{equation}
r\equiv \frac{3\epsilon_r}{4-\epsilon_r}.
\end{equation}
Furthermore, $S_\sigma$ can be expanded up to the second order as,
assuming that the curvaton has a quadratic potential, 
\begin{equation}
S_\sigma = 2 \frac{\delta \sigma_*}{ \sigma_*} - \left(\frac{\delta \sigma_*}{\sigma_*}\right)^2,
\end{equation}
where $\sigma_\ast = \bar{\sigma}_\ast + \delta \sigma_\ast$ is the value of the curvaton 
at the time of horizon exit and its fluctuations $\delta \sigma_\ast$ is assumed to be Gaussian.
We also treat $\zeta^{(\phi)}$ as a Gaussian field.

Although our setup here a bit
differs from one in Section \ref{sec:NG}, where
curvature and isocurvature perturbations are assumed to be 
generated from a single source, 
our result in Section \ref{sec:result} for correlated isocurvature
models can be still applicable as bispectra are generated from
the single source $S_\sigma$.
Then $\alpha$ and the parameters characterizing amplitudes of
bispectra, $f_{\rm NL}$ and $\alpha f_{\rm NL}^{\rm (ISO)}$, 
are given by 
\begin{eqnarray}
\alpha&=&\frac9A\left[\frac{\epsilon_{\rm CDM}}{r}-1\right]^2, 
\label{eq:alpha_cur} \\
f_{\rm NL}&=&\frac{5(1-r)(3+r)}{6rA^2}, 
\label{eq:fnl_cur} \\
\alpha f_{\rm NL}^{\rm (ISO)}&=&
\frac9{A^2}\left[
\frac{\epsilon_{\rm CDM}(1-\epsilon_{\rm CDM})}{2r^2}
-\frac{(1-r)(3+r)}{6r} \right], 
\label{eq:afnls_cur} 
\end{eqnarray}
where we defined a parameter $A$, which is the ratio of 
total curvature power to the contribution of the curvaton therein, i.e. 
\begin{equation}
A\equiv \frac{\langle \zeta^2\rangle}
{\langle (rS_\sigma/3)^2\rangle}.
\end{equation}

\begin{figure}
  \begin{center}
    \begin{tabular}{cc}
      \hspace{-10mm}\scalebox{1.}{\includegraphics{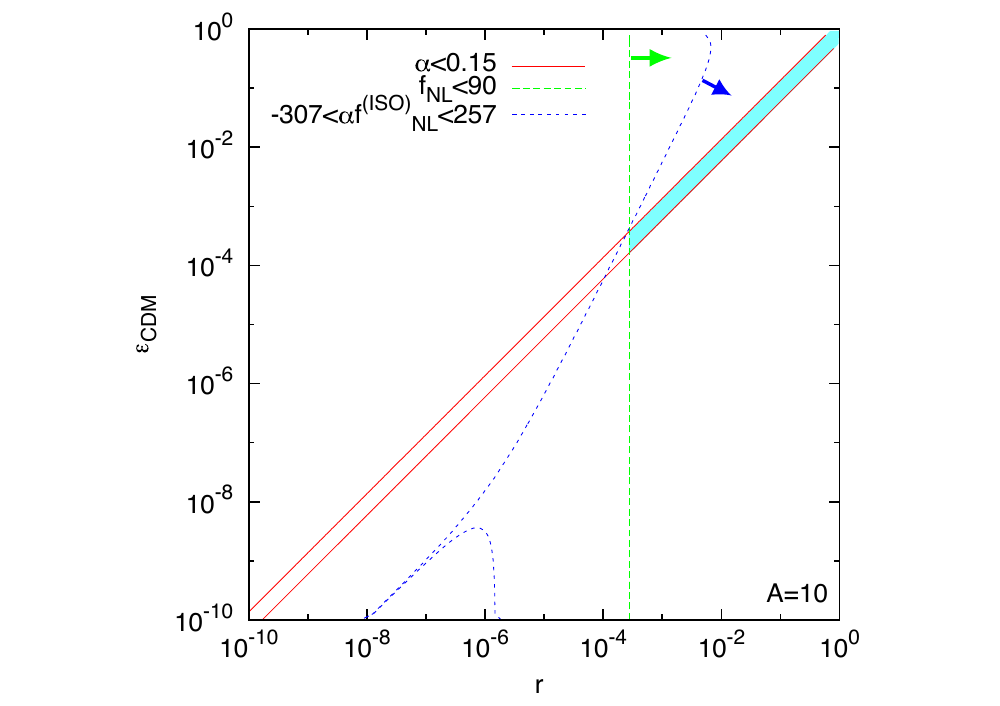}} &
      \hspace{-25mm}\scalebox{1.}{\includegraphics{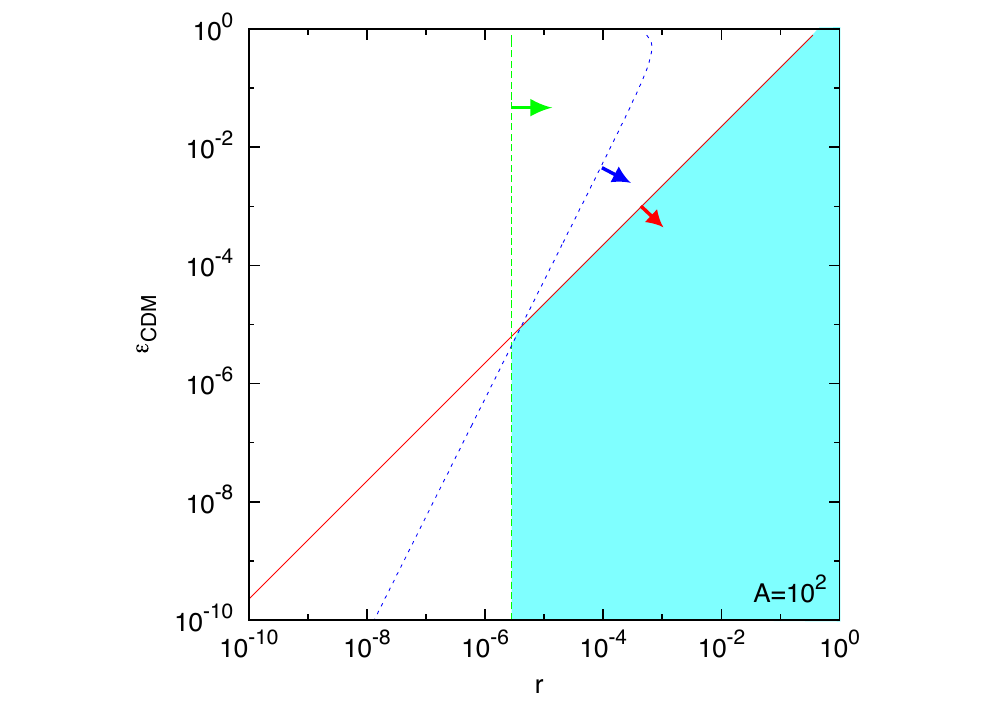}} \\
      \hspace{-10mm}\scalebox{1.}{\includegraphics{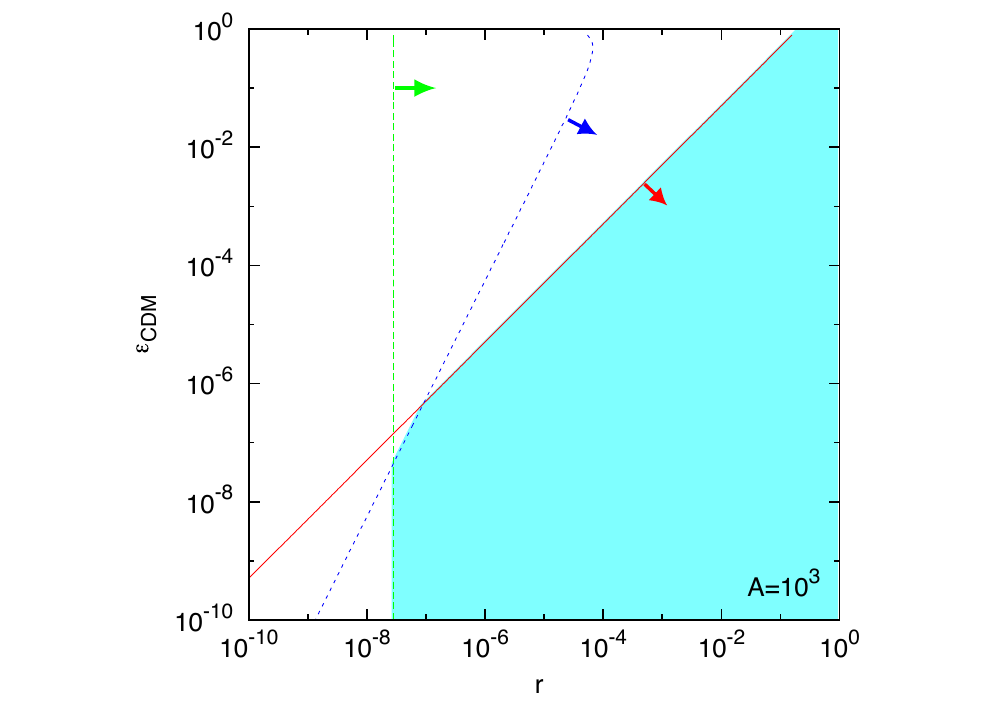}} &
      \hspace{-25mm}\scalebox{1.}{\includegraphics{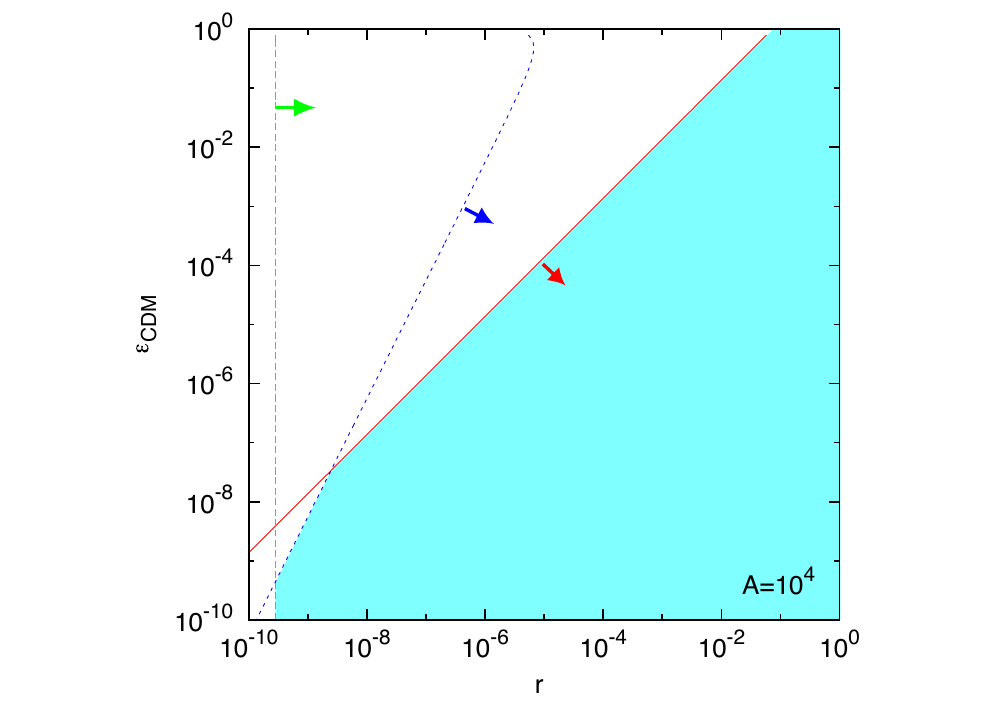}} 
    \end{tabular}
  \end{center}
  \caption{Constraints on curvaton scenario 
  in $r$-$\epsilon_{\rm CDM}$ plane with fixed values of $A$.
  From top left to bottom right panels, $A$ scales from 10 to $10^4$.
  An arrow indicates the direction of which region is allowed for each constraints. 
  Shaded regions are allowed  by the constraints $\alpha$, $f_{\rm NL}$ 
  and $\alpha f_{\rm NL}^{\rm (ISO)}$.
  For the case of $A=10$, the region allowed by the constraints from $\alpha$ is 
  the one between two red lines.
  See text for more details.
  }
  \label{fig:curvaton}
\end{figure}

Fig. \ref{fig:curvaton} shows constraints on $r$ and $\epsilon_{\rm CDM}$
at 2$\sigma$ level with several fixed values of $A$. Shaded regions
are allowed by the constraints $\alpha$, $f_{\rm NL}$ 
and $\alpha f_{\rm NL}^{\rm (ISO)}$.
Regarding the upper bound on $\alpha$,  when $A\gg1$, $S_{\rm CDM}$ is almost uncorrelated 
with the curvature perturbation since it almost originates to that from the inflaton. Thus Eq. \eqref{eq:CI_uncorr} is adopted.
As for bounds on $f_{\rm NL}$ and $f_{\rm NL}^{\rm (ISO)}$, 
we adopt those for correlated models with $n_{\rm adi}=n_{\rm iso}=1$
and template marginalization, 
$f_{\rm NL}<90$ and $-307<\alpha f_{\rm NL}^{\rm (ISO)}<257$.

When $A$ is as large as $\mathcal O(100)$, there is some parameter region 
where $\alpha f_{\rm NL}^{\rm (ISO)}$ can be observably large as 
$\alpha f_{\rm NL}^{\rm (ISO)}=\mathcal O(100)$ without producing too
large $\alpha$ and $f_{\rm NL}$. In such a case, the constraints on 
$\alpha f_{\rm NL}^{\rm (ISO)}$ becomes important.

\section{Conclusion} \label{sec:conclusion}

In this paper, we have investigated non-Gaussianties in isocurvature fluctuations
by extending our previous analysis for CDM uncorrelated mode \cite{Hikage:2012be} to 
a more general type. 
We derived optimal constraints on them using the CMB temperature maps
of the WMAP 7-year observation.
Specifically, we have studied correlated/uncorrelated CDM  
and neutrino density isocurvature modes along with the counterpart for 
the adiabatic one. Their constraints are summarized in Tables~\ref{tbl:uncorr} and \ref{tbl:corr} for 
uncorrelated and correlated modes, respectively. 
We have also presented joint analysis in which both the adiabatic and isocurvature fluctuations 
are assumed to exist and obtained 2D constraints as shown in Figs.~\ref{fig:uncorr} and \ref{fig:corr} 
for uncorrelated and correlated cases, respectively. 
We have checked our results by making  the Fisher matrix analysis  
adopting the specifications of WMAP7 experiments, which should 
give an idea how large the errors can be. 
Comparing with the results obtained by using actual CMB temperature maps,
we have found that they are almost the same in size, which shows the 
reasonability of our constraints. 

Our results indicate that both of the adiabatic and isocurvature perturbations, even if we 
consider various types for the latter, are consistent with 
Gaussian ones at around 2$\sigma$ level.  Although we have not detected any 
non-Gaussian signatures in those perturbations, their constraints may give 
interesting implications to some explicit models. As such an example, 
we have considered a curvaton scenario where correlated CDM fluctuations are generated. 
As mentioned in the text, isocurvature fluctuations have already been severely constrained by 
power spectrum measurements. However, we argued that, in some particular cases, 
our constraints for isocurvature non-Gaussianities can give a more stringent one.

In the near future, we are expecting to obtain much more accurate measurements of CMB from Planck. 
At this  era of precision cosmology, isocurvature fluctuations and their non-Gaussianities 
would provide fruitful information on the physics of the early Universe and the research along this line 
may also give important implications to particle physics as well.

\bigskip
\bigskip

\noindent 
\section*{Acknowledgment}

T.~S. would like to thank the Japan Society for the Promotion of Science for the financial report.
The authors acknowledge Kobayashi-Maskawa Institute for the Origin of
Particles and the Universe, Nagoya University for providing computing
resources  in conducting the research reported in this paper.
This work is supported by Grant-in-Aid for Scientific research from
the Ministry of Education, Science, Sports, and Culture (MEXT), Japan,
No.\ 14102004 (M.K.), No.\ 21111006 (M.K.), No.\ 23740195 (T.T.),  
No.\ 24740160 (C.H.) and also by World Premier International Research 
Center Initiative (WPI Initiative), MEXT, Japan. 
Some of the results in this paper have been derived using the HEALPix 
\cite{Gorski:2004by} package.
\appendix



\begin{thebibliography}{}

\bibitem{Bean:2006qz} 
  R.~Bean, J.~Dunkley and E.~Pierpaoli,
  Phys.\ Rev.\ D {\bf 74}, 063503 (2006)
  [astro-ph/0606685].

\bibitem{Moodley:2004nz} 
  K.~Moodley, M.~Bucher, J.~Dunkley, P.~G.~Ferreira and C.~Skordis,
  Phys.\ Rev.\ D {\bf 70}, 103520 (2004)
  [astro-ph/0407304].
  
\bibitem{Beltran:2004uv} 
  M.~Beltran, J.~Garcia-Bellido, J.~Lesgourgues and A.~Riazuelo,
  Phys.\ Rev.\ D {\bf 70}, 103530 (2004)
  [astro-ph/0409326].
    
\bibitem{Trotta:2006ww} 
  R.~Trotta,
  Mon.\ Not.\ Roy.\ Astron.\ Soc.\  {\bf 375}, L26 (2007)
  [astro-ph/0608116].
  
\bibitem{Keskitalo:2006qv} 
  R.~Keskitalo, H.~Kurki-Suonio, V.~Muhonen and J.~Valiviita,
  JCAP {\bf 0709}, 008 (2007)
  [astro-ph/0611917].
  
\bibitem{Kawasaki:2007mb} 
  M.~Kawasaki and T.~Sekiguchi,
  Prog.\ Theor.\ Phys.\  {\bf 120}, 995 (2008)
  [arXiv:0705.2853 [astro-ph]].
  
\bibitem{Sollom:2009vd} 
  I.~Sollom, A.~Challinor and M.~P.~Hobson,
  Phys.\ Rev.\ D {\bf 79}, 123521 (2009)
  [arXiv:0903.5257 [astro-ph.CO]].
    
\bibitem{Valiviita:2009bp} 
  J.~Valiviita and T.~Giannantonio,
  Phys.\ Rev.\ D {\bf 80}, 123516 (2009)
  [arXiv:0909.5190 [astro-ph.CO]].

\bibitem{Li:2010yb} 
  H.~Li, J.~Liu, J.~-Q.~Xia and Y.~-F.~Cai,
  Phys.\ Rev.\ D {\bf 83}, 123517 (2011)
  [arXiv:1012.2511 [astro-ph.CO]].

\bibitem{Valiviita:2012ub} 
  J.~Valiviita, M.~Savelainen, M.~Talvitie, H.~Kurki-Suonio and S.~Rusak,
  Astrophys.\ J.\  {\bf 753}, 151 (2012)
  [arXiv:1202.2852 [astro-ph.CO]].
  
\bibitem{Gordon:2009wx} 
  C.~Gordon and J.~R.~Pritchard,
  Phys.\ Rev.\ D {\bf 80}, 063535 (2009)
  [arXiv:0907.5400 [astro-ph.CO]].

\bibitem{Kawasaki:2011ze} 
  M.~Kawasaki, T.~Sekiguchi and T.~Takahashi,
  JCAP {\bf 1110}, 028 (2011)
  [arXiv:1104.5591 [astro-ph.CO]].

\bibitem{Hikage:2009rt} 
  C.~Hikage, D.~Munshi, A.~Heavens and P.~Coles,
  Mon.\ Not.\ Roy.\ Astron.\ Soc.\  {\bf 404}, 1505 (2010)
  [arXiv:0907.0261 [astro-ph.CO]].

\bibitem{Langlois:2011hn} 
  D.~Langlois and B.~van Tent,
  Class.\ Quant.\ Grav.\  {\bf 28}, 222001 (2011)
  [arXiv:1104.2567 [astro-ph.CO]].

\bibitem{Langlois:2012tm} 
  D.~Langlois and B.~van Tent,
  JCAP {\bf 1207}, 040 (2012)
  [arXiv:1204.5042 [astro-ph.CO]].

\bibitem{Kawakami:2012ke} 
  E.~Kawakami, M.~Kawasaki, K.~Miyamoto, K.~Nakayama and T.~Sekiguchi,
  JCAP {\bf 1207}, 037 (2012)
  [arXiv:1202.4890 [astro-ph.CO]].
  
\bibitem{Kawakami:2009iu} 
  E.~Kawakami, M.~Kawasaki, K.~Nakayama and F.~Takahashi,
  JCAP {\bf 0909}, 002 (2009)
  [arXiv:0905.1552 [astro-ph.CO]].
  
\bibitem{Langlois:2008vk} 
  D.~Langlois, F.~Vernizzi and D.~Wands,
  JCAP {\bf 0812}, 004 (2008)
  [arXiv:0809.4646 [astro-ph]].

\bibitem{Takahashi:2009cx}
  T.~Takahashi, M.~Yamaguchi and S.~Yokoyama,
  Phys.\ Rev.\  D {\bf 80}, 063524 (2009)
  [arXiv:0907.3052 [astro-ph.CO]].
  
\bibitem{Langlois:2011zz} 
  D.~Langlois and A.~Lepidi,
  JCAP {\bf 1101}, 008 (2011)
  [arXiv:1007.5498 [astro-ph.CO]].

\bibitem{Langlois:2010fe} 
  D.~Langlois and T.~Takahashi,
  JCAP {\bf 1102}, 020 (2011)
  [arXiv:1012.4885 [astro-ph.CO]].

\bibitem{Kawasaki:2008pa} 
  M.~Kawasaki, K.~Nakayama, T.~Sekiguchi, T.~Suyama and F.~Takahashi,
  JCAP {\bf 0901}, 042 (2009)
  [arXiv:0810.0208 [astro-ph]].

\bibitem{Kawasaki:2008sn} 
  M.~Kawasaki, K.~Nakayama, T.~Sekiguchi, T.~Suyama and F.~Takahashi,
  JCAP {\bf 0811}, 019 (2008)
  [arXiv:0808.0009 [astro-ph]].



\bibitem{Hikage:2008sk}
  C.~Hikage, K.~Koyama, T.~Matsubara, T.~Takahashi and M.~Yamaguchi,
  Mon.\ Not.\ Roy.\ Astron.\ Soc.\  {\bf 398}, 2188 (2009)
  [arXiv:0812.3500 [astro-ph]].
  
\bibitem{Hikage:2012be} 
  C.~Hikage, M.~Kawasaki, T.~Sekiguchi and T.~Takahashi,
  arXiv:1211.1095 [astro-ph.CO].








\bibitem{Axenides:1983hj}
  M.~Axenides, R.~H.~Brandenberger, M.~S.~Turner,
  Phys.\ Lett.\  {\bf B126}, 178 (1983).
  
\bibitem{Seckel:1985tj}
  D.~Seckel, M.~S.~Turner,
  Phys.\ Rev.\  {\bf D32}, 3178 (1985).
  
\bibitem{Linde:1985yf}
  A.~D.~Linde,
  Phys.\ Lett.\  {\bf B158}, 375-380 (1985).
  
\bibitem{Linde:1990yj}
  A.~D.~Linde, D.~H.~Lyth,
  Phys.\ Lett.\  {\bf B246}, 353-358 (1990).
  
\bibitem{Turner:1990uz}
  M.~S.~Turner, F.~Wilczek,
  Phys.\ Rev.\ Lett.\  {\bf 66}, 5 (1991).
   
\bibitem{Linde:1991km}
  A.~D.~Linde,
  Phys.\ Lett.\  {\bf B259}, 38 (1991).
  
\bibitem{Lyth:1991ub}
  D.~H.~Lyth,
  Phys.\ Rev.\  {\bf D45}, 3394 (1992).
  
\bibitem{Kasuya:2009up}
  S.~Kasuya, M.~Kawasaki,
  Phys.\ Rev.\  {\bf D80}, 023516 (2009)
  [arXiv:0904.3800 [astro-ph.CO]].



\bibitem{Enqvist:1998pf}
  K.~Enqvist and J.~McDonald,
  Phys.\ Rev.\ Lett.\  {\bf 83}, 2510 (1999)
  [arXiv:hep-ph/9811412].
  
\bibitem{Enqvist:1999hv}
  K.~Enqvist, J.~McDonald,
  Phys.\ Rev.\  {\bf D62}, 043502 (2000)
  [hep-ph/9912478].
  
\bibitem{Kawasaki:2001in}
  M.~Kawasaki and F.~Takahashi,
  Phys.\ Lett.\  B {\bf 516}, 388 (2001)
  [arXiv:hep-ph/0105134].
  
 
\bibitem{Enqvist:2001zp}
K.~Enqvist and M.~S.~Sloth,
Nucl.\ Phys.\ B {\bf 626}, 395 (2002)
[arXiv:hep-ph/0109214].

\bibitem{Lyth:2001nq}
D.~H.~Lyth and D.~Wands,
Phys.\ Lett.\ B {\bf 524}, 5 (2002)
[arXiv:hep-ph/0110002].

\bibitem{Moroi:2001ct}
T.~Moroi and T.~Takahashi,
Phys.\ Lett.\ B {\bf 522}, 215 (2001)
[Erratum-ibid.\ B {\bf 539}, 303 (2002)]
[arXiv:hep-ph/0110096].


\bibitem{Moroi:2008nn}
  T.~Moroi and T.~Takahashi,
  Phys.\ Lett.\  B {\bf 671}, 339 (2009)
  [arXiv:0810.0189 [hep-ph]].

\bibitem{Moroi:2002rd}
  T.~Moroi and T.~Takahashi,
  Phys.\ Rev.\  D {\bf 66}, 063501 (2002)
  [arXiv:hep-ph/0206026].

\bibitem{Moroi:2002vx} 
  T.~Moroi and H.~Murayama,
  Phys.\ Lett.\ B {\bf 553}, 126 (2003)
  [hep-ph/0211019].

\bibitem{Lyth:2002my}
  D.~H.~Lyth, C.~Ungarelli and D.~Wands,
  Phys.\ Rev.\  D {\bf 67}, 023503 (2003)
  [arXiv:astro-ph/0208055].


\bibitem{Lyth:2003ip}
  D.~H.~Lyth and D.~Wands,
  Phys.\ Rev.\  D {\bf 68}, 103516 (2003)
  [arXiv:astro-ph/0306500].


\bibitem{Hamaguchi:2003dc} 
  K.~Hamaguchi, M.~Kawasaki, T.~Moroi and F.~Takahashi,
  Phys.\ Rev.\ D {\bf 69}, 063504 (2004)
  [hep-ph/0308174].

\bibitem{Gordon:2003hw} 
  C.~Gordon and K.~A.~Malik,
  Phys.\ Rev.\ D {\bf 69}, 063508 (2004)
  [astro-ph/0311102].

\bibitem{Ikegami:2004ve} 
  M.~Ikegami and T.~Moroi,
  Phys.\ Rev.\ D {\bf 70}, 083515 (2004)
  [hep-ph/0404253].

\bibitem{Ferrer:2004nv} 
  F.~Ferrer, S.~Rasanen and J.~Valiviita,
  JCAP {\bf 0410}, 010 (2004)
  [astro-ph/0407300].

\bibitem{Beltran:2008ei}
  M.~Beltran,
  Phys.\ Rev.\  D {\bf 78}, 023530 (2008)
  [arXiv:0804.1097 [astro-ph]].
   
\bibitem{Lemoine:2009yu} 
  M.~Lemoine, J.~Martin and J.~'i.~Yokoyama,
  Europhys.\ Lett.\  {\bf 89}, 29001 (2010)
  [arXiv:0903.5428 [astro-ph.CO]].

\bibitem{Lemoine:2009is} 
  M.~Lemoine, J.~Martin and J.~'i.~Yokoyama,
  Phys.\ Rev.\ D {\bf 80}, 123514 (2009)
  [arXiv:0904.0126 [astro-ph.CO]].
  
\bibitem{Bucher:1999re} 
  M.~Bucher, K.~Moodley and N.~Turok,
  Phys.\ Rev.\ D {\bf 62}, 083508 (2000)
  [astro-ph/9904231].


\bibitem{Sasaki:1995aw} 
  M.~Sasaki and E.~D.~Stewart,
  Prog.\ Theor.\ Phys.\  {\bf 95}, 71 (1996)
  [astro-ph/9507001].

\bibitem{Lyth:2004gb} 
  D.~H.~Lyth, K.~A.~Malik and M.~Sasaki,
  JCAP {\bf 0505}, 004 (2005)
  [astro-ph/0411220].

\bibitem{Lyth:2005du} 
  D.~H.~Lyth and Y.~Rodriguez,
  Phys.\ Rev.\ D {\bf 71}, 123508 (2005)
  [astro-ph/0502578].

\bibitem{Lyth:2005fi} 
  D.~H.~Lyth and Y.~Rodriguez,
  Phys.\ Rev.\ Lett.\  {\bf 95}, 121302 (2005)
  [astro-ph/0504045].


\bibitem{Kawasaki:2008jy} 
  M.~Kawasaki, K.~Nakayama and F.~Takahashi,
  JCAP {\bf 0901}, 002 (2009)
  [arXiv:0809.2242 [hep-ph]].

\bibitem{Komatsu:2010fb} 
  E.~Komatsu {\it et al.}  [WMAP Collaboration],
  Astrophys.\ J.\ Suppl.\  {\bf 192}, 18 (2011)
  [arXiv:1001.4538 [astro-ph.CO]].

\bibitem{Dunkley:2010ge} 
  J.~Dunkley, R.~Hlozek, J.~Sievers, V.~Acquaviva, P.~A.~R.~Ade, P.~Aguirre, M.~Amiri and J.~W.~Appel {\it et al.},
  Astrophys.\ J.\  {\bf 739}, 52 (2011)
  [arXiv:1009.0866 [astro-ph.CO]].

\bibitem{Kawasaki:2011rc} 
  M.~Kawasaki, K.~Miyamoto, K.~Nakayama and T.~Sekiguchi,
  JCAP {\bf 1202}, 022 (2012)
  [arXiv:1107.4962 [astro-ph.CO]].

\bibitem{Komatsu:2001rj} 
  E.~Komatsu and D.~N.~Spergel,
  Phys.\ Rev.\ D {\bf 63}, 063002 (2001)
  [astro-ph/0005036].

\bibitem{Babich:2004yc} 
  D.~Babich and M.~Zaldarriaga,
  Phys.\ Rev.\ D {\bf 70}, 083005 (2004)
  [astro-ph/0408455].

\bibitem{Lewis:1999bs} 
  A.~Lewis, A.~Challinor and A.~Lasenby,
  Astrophys.\ J.\  {\bf 538}, 473 (2000)
  [astro-ph/9911177].

\bibitem{Knox:1995dq} 
  L.~Knox,
  Phys.\ Rev.\ D {\bf 52}, 4307 (1995)
  [astro-ph/9504054].

\bibitem{wmap}
  M.~Limon,
  ``Wilkinson Microwave Anisotropy Probe (WMAP): Seven Year Explanatory Supplement,"
  http://lambda.gsfc.nasa.gov/product/map/current/

\bibitem{Komatsu:2003iq} 
  E.~Komatsu, D.~N.~Spergel and B.~D.~Wandelt,
  Astrophys.\ J.\  {\bf 634}, 14 (2005)
  [astro-ph/0305189].

\bibitem{Yadav:2007rk} 
  A.~P.~S.~Yadav, E.~Komatsu and B.~D.~Wandelt,
  Astrophys.\ J.\  {\bf 664}, 680 (2007)
  [astro-ph/0701921].

\bibitem{Creminelli:2005hu} 
  P.~Creminelli, A.~Nicolis, L.~Senatore, M.~Tegmark and M.~Zaldarriaga,
  JCAP {\bf 0605}, 004 (2006)
  [astro-ph/0509029].
  
\bibitem{Yadav:2007ny} 
  A.~P.~S.~Yadav, E.~Komatsu, B.~D.~Wandelt, M.~Liguori, F.~K.~Hansen and S.~Matarrese,
  Astrophys.\ J.\  {\bf 678}, 578 (2008)
  [arXiv:0711.4933 [astro-ph]].

\bibitem{Smith:2009jr} 
  K.~M.~Smith, L.~Senatore and M.~Zaldarriaga,
  JCAP {\bf 0909}, 006 (2009)
  [arXiv:0901.2572 [astro-ph]].

\bibitem{Elsner:2009md} 
  F.~Elsner and B.~D.~Wandelt,
  Astrophys.\ J.\ Suppl.\  {\bf 184}, 264 (2009)
  [arXiv:0909.0009 [astro-ph.CO]].

\bibitem{Smith:2007rg} 
  K.~M.~Smith, O.~Zahn and O.~Dore,
  Phys.\ Rev.\ D {\bf 76}, 043510 (2007)
  [arXiv:0705.3980 [astro-ph]].

\bibitem{Jarosik:2010iu} 
  N.~Jarosik, C.~L.~Bennett, J.~Dunkley, B.~Gold, M.~R.~Greason, M.~Halpern, R.~S.~Hill and G.~Hinshaw {\it et al.},
  Astrophys.\ J.\ Suppl.\  {\bf 192}, 14 (2011)
  [arXiv:1001.4744 [astro-ph.CO]].

\bibitem{Gold:2010fm} 
  B.~Gold, N.~Odegard, J.~L.~Weiland, R.~S.~Hill, A.~Kogut, C.~L.~Bennett, G.~Hinshaw and J.~Dunkley {\it et al.},
  Astrophys.\ J.\ Suppl.\  {\bf 192}, 15 (2011)
  [arXiv:1001.4555 [astro-ph.GA]].
  
\bibitem{Gorski:2004by} 
  K.~M.~Gorski, E.~Hivon, A.~J.~Banday, B.~D.~Wandelt, F.~K.~Hansen, M.~Reinecke and M.~Bartelman,
  Astrophys.\ J.\  {\bf 622}, 759 (2005)
  [astro-ph/0409513].
    
\end{thebibliography}
\end{document}